\documentclass{article}
\usepackage{ijcai25}
\usepackage{graphicx}
\usepackage{subcaption}
\usepackage{tikz}
\usepackage{pgfplots}
\usepackage{amsmath}
\usepackage{amssymb}
\usetikzlibrary{positioning}
\usetikzlibrary{arrows.meta}
\usetikzlibrary{shapes.geometric}
\usetikzlibrary{backgrounds}
\usepackage{times}
\usepackage{soul}
\usepackage{booktabs}
\usepackage{url}
\usepackage[hidelinks]{hyperref}
\usepackage[utf8]{inputenc}
\usepackage{graphicx}
\usepackage{amsmath}
\usepackage{amsthm}
\usepackage{multirow}
\usepackage{booktabs}
\usepackage{booktabs}
\usepackage{algorithm}
\usepackage{algorithmic}
\pgfplotsset{compat=1.18}
\usepackage[switch]{lineno}

\title{RobSurv: Vector Quantization-Based Multi-Modal Learning for Robust Cancer Survival Prediction}
\author{
Submission Number: 6290
}

\author{
Aiman Farooq$^1$
\and
Azad Singh $^1$
Deepak Mishra$^1$\And
Santanu Chaudhury$^2$\\
\affiliations
$^1$Indian Institute of Technology Jodhpur\\
$^2$ Indian Institute of Technology Delhi\\
\emails
\{farooq.1,singh.63,dmishra\}@iitj.ac.in,
schaudhury@gmail.com.
}

\begin{document}

\maketitle

\begin{abstract}
    Cancer survival prediction using multi-modal medical imaging presents a critical challenge in oncology, mainly due to the vulnerability of deep learning models to noise and protocol variations across imaging centers. Current approaches struggle to extract consistent features from heterogeneous CT and PET images, limiting their clinical applicability. We address these challenges by introducing RobSurv, a robust deep-learning framework that leverages vector quantization for resilient multi-modal feature learning. The key innovation of our approach lies in its dual-path architecture: one path maps continuous imaging features to learned discrete codebooks for noise-resistant representation, while the parallel path preserves fine-grained details through continuous feature processing. This dual representation is integrated through a novel patch-wise fusion mechanism that maintains local spatial relationships while capturing global context via Transformer-based processing. In extensive evaluations across three diverse datasets (HECKTOR, H\&N1, and NSCLC Radiogenomics), RobSurv demonstrates superior performance, achieving concordance index of 0.771, 0.742, and 0.734 respectively - significantly outperforming existing methods. Most notably, our model maintains robust performance even under severe noise conditions, with performance degradation of only 3.8-4.5\% compared to 8-12\% in baseline methods. These results, combined with strong generalization across different cancer types and imaging protocols, establish RobSurv as a promising solution for reliable clinical prognosis that can enhance treatment planning and patient care.

\end{abstract}

\section{Introduction}

Cancer survival prediction is a critical challenge in healthcare, with significant implications for treatment planning and patient outcomes ~\cite{mariotto2014cancer}. Accurate survival time prediction enables clinicians to make informed decisions about treatment strategies, optimize resource allocation, and improve patient care through personalized medicine approaches. The availability of multi-modal data, particularly medical imaging (CT, PET scans), genomic information, and clinical records, have created new opportunities for more accurate predictions while introducing unique analytical challenges ~\cite{debiase2024deep,saeed2024survrnc}. Recent advances in medical imaging technologies have made it possible to capture detailed anatomical and functional information, potentially offering new insights into disease progression and patient outcomes. Integrating multiple imaging modalities, particularly CT and PET scans, presents opportunities and allows us to capture and leverage complementary contextual cues across diverse modalities. While multiple data modalities enhance predictions, noise within datasets poses significant challenges, obscuring critical patterns necessary for accurate predictions. Despite the challenges, the available modalities still contain valuable information, making it essential to develop robust methods that mitigate noise and maximize the utility of noisy data in survival prediction models.

Survival prediction has evolved significantly from traditional statistical approaches to sophisticated deep-learning (DL) methods. Early work primarily relied on the Cox proportional hazards model ~\cite{cox1972regression}, and random survival forests ~\cite{he2022artificial}, which provided interpretable results and gained widespread clinical adoption. Convolutional Neural Network (CNN)-based models ~\cite{katzman2018deepsurv,lian2022imaging} surpass traditional handcrafted methods by extracting imaging features for constructing effective survival prediction models. However, these methods face challenges in capturing complex non-linear relationships within high-dimensional multi-modal data, oversimplifying the intricate biological processes underlying cancer progression, effective mechanisms for integrating heterogeneous data types, and vulnerability to noise. Recent works, such as ~\cite{farooq2024survival,ma2024deep,debiase2024deep}, have demonstrated the potential of integrating complementary information from diverse modalities to enhance survival prediction, achieving superior performance across various cancer types.
~\cite{li2020robust} presents a robust feature extraction method to learn noise-invariant representations and adaptive fusion strategies. ~\cite{ma2024deep} developed a fusion strategy that dynamically adjusts the contribution of each modality based on data quality and reliability.

Existing approaches to survival prediction face several fundamental challenges that limit their clinical applicability. Although models demonstrate high accuracy on clean datasets, they often struggle in real-world clinical environments where data quality is variable and noise is inevitable. While traditional denoising methods can improve image quality, they frequently introduce subtle artifacts that compromise downstream analysis, highlighting the need for inherently robust modeling approaches. Integrating multi-modal data to leverage complementary information poses significant challenges. While PET scans provide functional insights through metabolic patterns and CT scans offer detailed anatomical structures, combining these diverse modalities is technically complex. The differences in data distributions, spatial resolutions, and noise characteristics across modalities make alignment and fusion particularly difficult, especially when retaining modality-specific information. Additionally, the three-dimensional nature of medical imaging data increases the complexity, as models must preserve spatial relationships while effectively capturing meaningful features. These challenges underscore the critical need for approaches that can maintain robust performance across diverse clinical settings while effectively leveraging the complementary strengths of multiple imaging modalities.

To address the key challenges in multi-modal cancer survival prediction, we propose a novel approach that effectively integrates CT and PET imaging modalities while ensuring robustness against noise and data perturbations. Our method introduces several key innovations that advance the state of the art:

\begin{enumerate}
    \item We are the first to incorporate vector quantization in cancer survival prediction, demonstrating its ability to enhance model robustness against noise and protocol variations. Our DualVQ framework generates complementary discrete and continuous features from CT-PET pairs. Discrete tokens capture stable anatomical and metabolic patterns, while continuous features retain fine-grained intensity variations critical for prognosis.
    
    \item Our DualPathFuse architecture processes discrete and continuous features in separate pathways to maximize their complementary strengths. The discrete pathway utilizes patch-wise bidirectional attention to model cross-modal relationships from CT to PET and PET to CT, while the continuous pathway employs channel-spatial attention to identify and localize survival-critical markers. This allows the model to effectively capture correlations between anatomical structures and metabolic activities while preserving modality-specific characteristics.
    
    \item Our solution is evaluated across multi-modal cancer survival datasets under clean and noisy clinical conditions. \textit{RobSurv} demonstrates robust prognostic accuracy, achieving state-of-the-art concordance indices and maintaining performance stability with minor degradation under severe noise—outperforming existing methods by 4–8\%, enabling reliable risk stratification $(\rho \leq 0.05)$ even in noisy imaging conditions.
    
\end{enumerate}


We evaluate our approach through extensive experiments on multiple cancer datasets, comparing it against state-of-the-art methods under realistic noise conditions. Our results demonstrate significant improvements in prediction accuracy and robustness, with our method maintaining high performance even under noise levels that cause substantial degradation in baseline approaches.

\begin{figure*}[t]
    \centering
    \includegraphics[width=\textwidth]{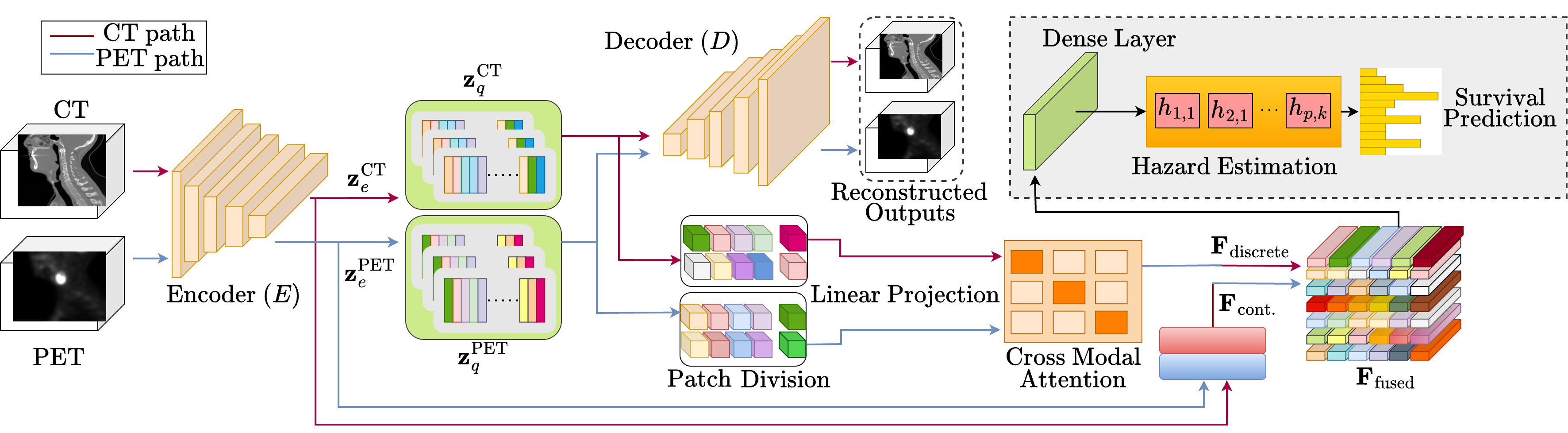}
    \caption{Architecture of the proposed \textit{RobSurv} architecture for robust survival prediction, consisting of three key components: Dual Vector Quantization (DualVQ) for parallel CT-PET processing, DualPatchFuse for cross-modal attention and fusion, and a hybrid survival prediction network for patient outcome estimation with competing risk analysis.}

    \label{fig:fig1}
\end{figure*}
\section{Related Work}
\subsection{Survival Prediction}

Cancer survival prediction has significantly transformed with DL approaches, evolving from traditional statistical methods to sophisticated multi-modal architectures. Early DL approaches focused on extracting meaningful features from individual imaging modalities. ~\cite{haarburger2019image} demonstrated the viability of using convolutional neural networks for extracting prognostic features from 3D medical images in lung cancer prediction. Their work addressed computational challenges through efficient batch processing while simplifying the complex survival analysis task to median survival classification. 

Integrating multiple imaging modalities has emerged as a crucial direction for improving prediction accuracy. ~\cite{saeed2022tmss} process CT and PET modalities independently through separate CNN streams before prediction and achieve a concordance index of 0.72 for head and neck cancer. However, this independent processing limited the model's capture of cross-modal interactions.  ~\cite{meng2023merging} addressed this limitation through their Hybrid Parallel Cross-Attention (HPCA) block, which enabled simultaneous CT and PET data processing using parallel convolutional layers and cross-attention transformers. Their addition of a Region-specific Attention Gate (RAG) highlighted the importance of spatial awareness in feature fusion. Building upon this foundation, ~\cite{saeed2024survrnc} introduced the Survival Rank-N-Contrast (SurvRNC) method, significantly advancing feature learning through a novel loss function that enforces ordinal relationships in learned representations. Their approach uniquely addresses the challenge of censored data.
~\cite{dao2022survival} developed a parallel transformer mechanism that effectively captures global context through multi-scale feature processing while preserving spatial relationships via positional encoding. ~\cite{wang2022novel} built upon this foundation through hierarchical attention mechanisms that dynamically weight different imaging features based on their prognostic value. These transformer-based approaches demonstrated the potential of attention mechanisms in capturing long-range dependencies and complex cross-modal relationships, though they typically require substantial computational resources.

Recent advances have focused on more sophisticated integration strategies. ~\cite{debiase2024deep} demonstrated the value of uncertainty consideration by incorporating probability maps of primary tumors alongside traditional imaging data. ~\cite{farooq2024survival} introduced a comprehensive approach that integrates CT, PET, and genomic data through self-supervised learning and explicitly models cross-patient relationships.

\subsection{Feature Quantization in Medical Imaging}

Feature quantization has emerged as a powerful technique in medical image analysis, offering robustness, interpretability, and computational efficiency benefits. Discretizing continuous features into learned codebooks has shown particular promise in handling the complex nature of medical imaging data. Recent advances in self-supervised learning have laid the groundwork for effective feature quantization approaches. Methods like SimCLR ~\cite{chen2020simple}, MoCo ~\cite{he2020momentum}, and BYOL ~\cite{grill2020bootstrap} have demonstrated the importance of learning robust representations, particularly in medical imaging contexts. These approaches have been adapted specifically for medical applications, as seen in MoCo-CXR ~\cite{sriram2021covid} for chest X-rays and MICLe ~\cite{azizi2021big} for multi-instance learning. Vector quantization, particularly through approaches like VQ-GAN ~\cite{esser2021taming}, has shown promising results in medical image analysis. ~\cite{zhang2023vector} uses a codebook for learning an effective discrete representation to achieve cross-modality image generation capacity. ~\cite{gorade2024synergynet} leverages the strengths of both continuous latent space (CLS) and discrete latent space (DLS) models to capture fine-grained details and coarse-grained structural information. CoBooM framework ~\cite{singh2024coboom} demonstrates an effective approach to capture and utilize anatomical similarities in medical images, leading to more informative and robust representations. This approach is particularly relevant for tasks requiring fine-grained feature discrimination, such as survival prediction.

However, existing approaches to quantization in medical imaging analysis have primarily focused on single-modality scenarios and tasks such as segmentation and classification. The challenge of effectively quantifying features from multiple imaging modalities, such as CT and PET, while preserving complementary information remains largely unexplored. While previous works have demonstrated the benefits of quantization for various medical tasks ~\cite{khader2023denoising,zhou2024conditional}, the specific requirements of survival prediction, particularly capturing temporal dependencies and prognostic factors, present unique challenges. Most existing approaches either process modalities independently or use relatively simple fusion strategies. The robustness of these models in terms of noise and variations in image quality has not been thoroughly addressed. Our work addresses these limitations through a novel quantization-based approach combined with attention mechanisms, enabling a more robust and effective fusion of CT and PET imaging data while maintaining the benefits of discrete representation learning.

\section{Methodology}
\paragraph{Problem Description.}The dataset $\mathcal{D} = \{P_1, P_2, \ldots, P_N\}$ consists of $N$ patients, where each patient $P_i$ is characterized by a pair of imaging modalities: CT scans $X_i^{\text{CT}} \in {R}^{H_{\text{CT}} \times W_{\text{CT}} \times D_{\text{CT}}}$ and PET scans $X_i^{\text{PET}} \in {R}^{H_{\text{PET}} \times W_{\text{PET}} \times D_{\text{PET}}}$. The network learns a feature representation through an end-to-end DL architecture, mapping the multi-modal medical images to hazard functions $h_{p,k}^{(i)}$ that estimate the probability of experiencing risk $k$ within time interval $p$, and corresponding cumulative incidence functions that capture the survival dynamics across multiple competing risks.

As shown in Figure ~\ref{fig:fig1}, our model comprises of three primary modules: Dual Vector Quantization (DualVQ), DualPatchFuse, and survival prediction. CT and PET volumes are initially processed through a 3D feature extractor, generating high-dimensional representations. These representations undergo parallel processing: one pathway preserves the original continuous features while the other transforms them through a vector quantization mechanism. The continuous and discrete features are then subjected to a transformer-based patch-wise fusion strategy, which captures cross-modal interactions and extracts global contextual representations. Combining the local patch-level and global transformers, a hybrid attention mechanism enables comprehensive feature learning. The fused representations are subsequently fed into a modified survival prediction network~\cite{lee2018deephit}, which learns the joint distribution of survival times and competing risks.

\subsection{Dual Vector Quantization (DualVQ)}
Our DualVQ framework extends the 3D VQ-GAN ~\cite{zhou2024conditional} architecture to simultaneously handle dual modalities (CT and PET) through parallel processing streams. Vector quantization is particularly beneficial for medical imaging as it helps capture discrete, anatomically meaningful features while reducing image noise and redundancy. The proposed framework maintains separate encoders, codebooks, and decoders for each modality while sharing the same architectural design principles. The encoder $E$ and decoder $D$ follow identical architectures with five 3D convolutional layers and five residual blocks. $E$ maps input volumes $X^{\text{CT}}, X^{\text{PET}} \in {R}^{ 128 \times 128 \times 128 }$ to latent representations $\mathbf{z}^{\text{CT}}_e, \mathbf{z}^{\text{PET}}_e \in {R}^{B \times 512 \times 8\times 8\times 8}$, where $B$ is the batch size. In the quantization step, $\mathbf{z}^{\text{CT}}_e, \mathbf{z}^{\text{PET}}_e$ undergo discretization through vector quantization. Formally, for each modality $m \in \{\text{CT}, \text{PET}\}$, we train a learnable codebook $\mathcal{C}_m$ where $\mathcal{C}_m = \{{c}_{b=1,...,1024}\}$ that transforms the encoded feature into discrete latent tokens. The process maps each spatial location's feature vector to its nearest codebook entry using Euclidean distance, producing $8 \times 8 \times 8$ discrete tokens. We maintain separate codebooks for each modality to preserve modality-specific characteristics. Each modality captures fundamentally different biological information - CT represents anatomical structure, while PET captures metabolic activity. This separation ensures that the distinct characteristics of each modality are preserved and effectively encoded in their respective discrete representation spaces. The quantization mechanism for each modality $m \in \{\text{CT}, \text{PET}\}$ is formally defined as:
\begin{equation}
\mathbf{z}^m_q = \arg\min_{{c}_k \in \mathcal{C}_m} |{\mathbf{z}}^m_e - {c}_k|_2
\end{equation}
where $\mathbf{z}^m_q$ is the discrete representation for modality $m$.

The total quantization loss $\mathcal{L}_q$ combines losses from both modalities:
\begin{equation}
\mathcal{L}_q = \sum_{m \in \{\text{CT}, \text{PET}\}} \mathcal{L}^m_q
\end{equation}
where for each modality, the quantization loss comprises of:
\begin{equation}
\mathcal{L}^m_q = \mathcal{L}^m_{{cb}} + \alpha_1 \cdot \mathcal{L}^m_{{ce}} + \alpha_2 \cdot \mathcal{L}^m_{{recon}}
\end{equation}

\begin{equation}
    \mathcal{L}^m_{{cb}} = |\text{SG}[{\mathbf{z}}^m_e] - {c}_k|_2^2
    \label{eqn:codebook}
\end{equation}

\begin{equation}
    \mathcal{L}^m_{{ce}} = |{\mathbf{z}}^m_e - \text{SG}[{c}_k]|_2^2
        \label{eqn:commit}
\end{equation}
\begin{equation}
    \mathcal{L}^m_{{recon}} = |\mathbf{x}_m - \hat{\mathbf{x}}_m|_2^2
    \label{eqn:recon}
\end{equation}
where, ~\eqref{eqn:codebook} referes to the codebook, ~\eqref{eqn:commit} commitment, and ~\eqref{eqn:recon} reconstructions losses, respectively, $\text{SG}$ denotes the stop-gradient operator, $\alpha_1$ and $\alpha_2$ are balancing hyperparameters. The codebook loss updates the codebook vectors to match the encoder outputs better, while the commitment loss ensures encoded features remain close to their assigned codebook vectors. The reconstruction loss helps maintain the fidelity of the reconstructed features for each modality. This dual-stream quantization approach allows each modality to develop its specialized discrete representation space while preserving modality-specific characteristics.

The discrete feature vectors for each modality are fed into their respective decoders $D$ to produce reconstructed volumes. For each modality, the discrete features $\mathbf{z}_q^m \in {R}^{512 \times 8 \times 8 \times 8}$ are processed through a series of upsampling blocks to restore the original spatial dimensions. The DualVQ module outputs both discrete tokens $(\mathbf{z}_q^{{\text{CT}}}, \mathbf{z}_q^{\text{PET}})$ and continuous features $(\mathbf{z}_e^{{\text{CT}}},\mathbf{z}_e^{\text{PET}})$ for each modality. These parallel representations serve complementary purposes: the discrete tokens capture robust anatomical patterns and key structural information through quantization, while the continuous features preserve fine-grained details and subtle intensity variations. Both representations are fed into our DualPatchFuse module for effective cross-modal integration.

\subsection{DualPatchFuse: A Patch-wise Cross-Modal Fusion Strategy}
Our fusion strategy employs a dual-path approach to leverage discrete and continuous features, combining their strengths. Cross-attention mechanisms are applied to quantized features, which are discretized into a finite codebook for structured, noise-resistant representations of anatomical patterns. This design ensures robust learning of stable cross-modal relationships by mitigating the influence of noise and spurious correlations, a common challenge in continuous representations.

For the CT and PET discrete feature vectors, $\mathbf{z}_q^{\text{CT}}$ and $\mathbf{z}_q^{\text{PET}}$, we perform patch-wise cross-modal fusion to generate a single output. Each vector is divided into smaller, non-overlapping 3D patches of size $2 \times 2 \times 2$. This results in $P = (8/2) \times (8/2) \times(8/2) =64 $ patches. We adopt a patch-based approach as medical features often have local spatial dependencies. The $2\times 2 \times 2$ patch size balances local feature capture with computational efficiency, allowing the model to learn meaningful regional patterns while maintaining manageable computational complexity. Each patch is then flattened and projected into a lower-dimensional embedding space using a learnable linear projection, producing embeddings $\mathbf{E}_{\text{CT}} \in {R}^{B \times P \times d_{\text{model}}}$ and $\mathbf{E}_{\text{PET}} \in {R}^{B \times P \times d_{\text{model}}}$, where $d_{\text{model}}$ is the embedding dimension. Positional encodings are added to these embeddings to retain spatial relationships.
The cross-modal interaction is modeled using bidirectional patch-wise cross-attention. The query, key, and value projections are computed for both modalities. $\mathbf{Q}_m$, $\mathbf{K}_m$, and $\mathbf{V}_m$ represent the modality-specific query, key, and value projection matrices, respectively. The bidirectional attention weights are computed using scaled dot-product attention:


 
\begin{equation}
\begin{split}
    \mathbf{A}_{\text{CT}\rightarrow\text{PET}}(i, j) &= \textbf{S} \left( \frac{\mathbf{Q}_{\text{CT}}[i] \cdot \mathbf{K}_{\text{PET}}[j]^T}{\sqrt{d_k}} \right) \cdot \mathbf{V}_{\text{PET}}[j] \\
    \mathbf{A}_{\text{PET}\rightarrow\text{CT}}(i, j) &= \textbf{S} \left( \frac{\mathbf{Q}_{\text{PET}}[i] \cdot \mathbf{K}_{\text{CT}}[j]^T}{\sqrt{d_k}} \right) \cdot \mathbf{V}_{\text{CT}}[j]
\end{split}
\end{equation}
where $d_k$ is the query/key dimension and the softmax function is denoted as \textbf{S}. These bidirectional attention weights enable the model to learn the mutual relationships between CT and PET features. The CT $\rightarrow$ PET attention allows the network to identify relevant metabolic activities for each anatomical region, while the PET $\rightarrow$ CT attention helps focus on structural regions corresponding to significant metabolic patterns. This results in two complementary fused embeddings $\mathbf{A}_{\text{CT}\rightarrow\text{PET}}$ and $\mathbf{A}_{\text{PET}\rightarrow\text{CT}} \in {R}^{B \times P \times d_k}$.

The discrete fused representation ($\mathbf{F}_{\text{discrete}}$) is obtained by combining both attention outputs and original embeddings with learnable modality-specific weights:
\begin{equation}
    \mathbf{F}_{\text{discrete}} = \beta_1(\mathbf{A}_{\text{CT}\rightarrow\text{PET}} + \mathbf{E}_{\text{CT}}) + \beta_2(\mathbf{A}_{\text{PET}\rightarrow\text{CT}} + \mathbf{E}_{\text{PET}})
\end{equation}
where $\beta_1, \beta_2 \in {R}$ are learnable parameters. Finally, the fused patch embeddings are reshaped back to the original 3D spatial dimensions, resulting in $\mathbf{V}_{\text{fused}}$. The fusion process is guided by two losses: a cross-modal alignment loss $\mathcal{L}_{{align}}$, which encourages bidirectional consistency between corresponding CT and PET features, and an information preservation loss $\mathcal{L}_{{preserve}}$, which ensures that the fused representation retains important information from both modalities equally.

\begin{equation}
\mathcal{L}_{{align}} = -\sum_{i=1}^N \frac{\mathbf{Q}_{\text{CT}}[i] \cdot \mathbf{K}_{\text{PET}}[i]}{|\mathbf{Q}_{\text{CT}}[i]| |\mathbf{K}_{\text{PET}}[i]|}
\end{equation}

\begin{equation}
\mathcal{L}_{{preserve}} = |\mathbf{F}_{\text{discrete}} - \mathbf{E}_{\text{CT}}|_2 + \beta_3|\mathbf{F}_{\text{discrete}} - \mathbf{E}_{\text{PET}}|_2
\end{equation}

\begin{equation}
   \mathcal{L}_{fusion}=\mathcal{L}_{{align}}+\mathcal{L}_{{preserve}}
\end{equation}
where $\beta_3$ is a balancing hyperparameter.

We use a two-step attention process for the continuous feature vectors following ~\cite{woo2018cbam}. First, channel attention identifies which features are most important for predicting survival. Then, spatial attention pinpoints where these critical features are located in the CT and PET scans. This dual attention mechanism helps the network focus on the most relevant information from both imaging modalities and generate the fused continuous representation $(\mathbf{F}_{\text{cont.}})$.

The fusion process results in two key representations: the patch-wise fused features from discrete tokens $(\mathbf{F}_{\text{discrete}})$ and the attention-weighted continuous features $(\mathbf{F}_{\text{cont.}})$. These complementary representations capture robust structural and anatomical patterns, intensity variations, and fine-grained details. The fused representations are then concatenated and processed through a fully connected layer to create a comprehensive feature vector $(\mathbf{F}_{\text{fused}})$that captures both local and global characteristics of the CT-PET pair. This rich representation serves as input to our hybrid survival prediction network.

\subsection{Hybrid Survival Prediction}

Our hybrid survival prediction framework extends traditional discrete-time analysis by incorporating competing risks and multi-modal data integration. We minimize a total survival loss $\mathcal{L}_{surv}$, combining a likelihood loss $\mathcal{L}_{likelihood}$ and a ranking loss $\mathcal{L}_{ranking}$. The likelihood loss is computed as:
\begin{align}
\mathcal{L}_{{likelihood}} = -\sum_{i=1}^N \Bigg[ \sum_{p=1}^P \sum_{k=1}^K {1}(t^{(i)} = p, e^{(i)} = k) \log(h_{p,k}^{(i)}) \nonumber \\
+ \sum_{p=1}^{t^{(i)} - 1} \log\left( 1 - \sum_{k=1}^K h_{p,k}^{(i)} \right) \Bigg]
\end{align}
where, ${1}(t^{(i)} = p, e^{(i)} = k)$ is an indicator function for observed events. The first term encourages accurate hazard predictions for observed events, while the second term promotes higher survival probabilities for censored subjects. The ranking loss $\mathcal{L}_{ranking}$ ensures proper risk stratification through pairwise comparisons :

\begin{equation}
\mathcal{L}_{{ranking}} = \sum_{k=1}^K \lambda_k \sum_{i\neq j} A_{k,i,j} \cdot \eta\left(F_{k,t^{(i)}}^{(i)}, F_{k,t^{(j)}}^{(j)}\right),
\end{equation}
where the coefficients $\lambda_k$ are chosen to trade off ranking losses of the $k-$th competing event, $A_{k,i,j}$ indicates valid comparison pairs for risk $k$, and $\eta(\cdot,\cdot)$ is a convex loss function. Further details are provided in the supplementary.

\subsubsection{Final Objective}
The objective function integrates quantization, cross-modal fusion, and survival prediction components.
\begin{equation}
\mathcal{L}_{{total}} = \gamma_1 \cdot \mathcal{L}_q + \gamma_2 \cdot \mathcal{L}_{fusion} + \gamma_3 \cdot \mathcal{L}_{{surv}}
\end{equation}
where $\gamma$ parameters balance the contribution of quantization and survival objectives.
\begin{table*}[htbp]
\centering
\caption{Evaluation of model performance (\(C_{td}\)-index) across various datasets under normal and noisy conditions, with 50\% of the samples containing noise. The best results are in bold, while the second-best are underlined.}
\label{tab:table1}
\begin{tabular}{lcccccc}
\toprule
\multirow{2}{*}{Model} & \multicolumn{3}{c}{\(C_{td}\)-index for clean samples} & \multicolumn{3}{c}{\(C_{td}\)-index for noisy samples} \\
\cmidrule(lr){2-4} \cmidrule(lr){5-7}
& NSCLC & HECKTOR & H\&N1 & NSCLC & HECKTOR & H\&N1 \\ 
\midrule

CoxPH & 0.641{$\pm$ 0.01} & 0.667 $\pm$ 0.07 & 0.632 $\pm$ 0.06 & 0.480 $\pm$ 0.03 & 0.512$\pm$ 0.09 & 0.509 $\pm$ 0.08 \\
DeepSurv & 0.672 $\pm$ 0.04 & 0.671 $\pm$ 0.03 & 0.668 $\pm$ 0.04 & 0.549 $\pm$ 0.05  &  0.560 $\pm$ 0.05 & 0.558 $\pm$ 0.03 \\
Multimodal Dropout & 0.689 $\pm$ 0.06 & 0.680 $\pm$ 0.66 & 0.700 $\pm$ 0.01 & 0.581 $\pm$ 0.07 & 0.583 $\pm$ 0.67 & 0.601 $\pm$ 0.02\\

DeepMTS & 0.715 $\pm$ 0.05 & 0.675 $\pm$ 0.02 & 0.724 $\pm$ 0.03 & 0.617 $\pm$ 0.06 & 0.591 $\pm$ 0.03 & 0.624 $\pm$ 0.04\\
 
XSurv & \underline{0.721 $\pm$ 0.06} & 0.710 $\pm$ 0.02 & 0.722 $\pm$ 0.01 & 0.652 $\pm$ 0.07 & 0.641 $\pm$ 0.03 & 0.614 $\pm$ 0.02  \\ 
MMRL & 0.672 $\pm$ 0.20 &  0.715 $\pm$ 0.03& 0.692 $\pm$ 0.05 & 0.608 $\pm$ 0.21 &  0.617 $\pm$ 0.04 &  0.595 $\pm$ 0.06 \\ 
SurvRNC & 0.711 $\pm$ 0.12 & 0.720 $\pm$ 0.05 & 0.681 $\pm$ 0.01 & 0.653 $\pm$ 0.13 & 0.669 $\pm$ 0.06 & 0.642 $\pm$ 0.02\\ 
TMSS & 0.708 $\pm$ 0.05 &  \underline{0.751 $\pm$  0.17} & \underline{0.725 $\pm$ 0.02} & \underline{0.661 $\pm$ 0.06} & \underline{0.712 $\pm$ 0.07} &  \underline{0.689 $\pm$ 0.03 }\\ 
RobSurv & \textbf{0.734 $\pm$ 0.02} & \textbf{0.771 $\pm$ 0.13} & \textbf{0.742 $\pm$ 0.02} & \textbf{0.701 $\pm$ 0.03} & \textbf{0.742 $\pm$ 0.14} & \textbf{ 0.702 $\pm$ 0.03} \\ 
\bottomrule
\end{tabular}
\end{table*}
\section{Experiment Setup}

\subsection{Datasets}
We utilized three cancer datasets for this study: the HEad and neCK TumOR segmentation and outcome prediction (HECKTOR) dataset ~\cite{oreiller2022head}, the HEAD-NECK-RADIOMICS-HN1 collection ~\cite{wee2019}, and NSCLC Radiogenomics ~\cite{bakr2017nsclc,bakr2018radiogenomic} dataset. The HECKTOR dataset includes PET and CT scans from 488 patients with tumor segmentation masks from seven different centers and Recurrence-Free Survival (RFS) information, including time-to-event data and censoring status. The HEAD-NECK-RADIOMICS-HN1 (H\&N1) collection comprises imaging and clinical data for 137 head-and-neck squamous cell carcinoma (HNSCC) patients, while the NSCLC Radiogenomics (NSCLC) dataset contains data from 211 patients. 

\subsubsection{Noisy Data} To evaluate model robustness, we augment our dataset by introducing synthetic noise to both modalities. CT images were corrupted with Gaussian noise, while PET images were degraded with Poisson noise at varying count levels. For CT images, we simulated acquisition noise by adding zero-mean Gaussian noise with varying standard deviations $(\sigma = [0.01,0.05,0.1])$ relative to the image intensity range. PET images were corrupted with low, medium, and high Poisson noise to mimic the statistical nature of photon counting noise inherent in nuclear imaging. Detailed dataset descriptions and noise injection protocols are provided in the supplementary.

\subsection{Implementation Details \& Model Evaluation }
We train the models on NVIDIA A100 Tensor Core and RTX 3090 GPU using the PyTorch framework. For backbone encoders, we use ResNet18 ~\cite{he2016deep}. For the DualVQ loss components ($\mathcal{L}_q$), we set $\alpha_1 = 0.25$ for the commitment loss weight and $\alpha_2 = 1.0$ for the reconstruction loss weight. In the cross-modal fusion module, we use an embedding dimension $d_{\text{model}} = 256$ and query/key dimension $d_k = 64$, with initial modality fusion weights $\beta_1 = \beta_2 = 0.5$ and preservation loss weight $\beta_3 = 0.5$. The final objective weights are set as $\gamma_1 = 1.0$, $\gamma_2 = 0.5$, and  $\gamma_3 = 2.0$. The training configuration uses a batch size of 2, a learning rate of $1 \times 10^{-4}$ with Adam optimizer. 

We adopt a v-fold cross-validation setup for both training and evaluation. We compute the test set's average performance in each run by leveraging the best-performing models' results. We use the time-dependent concordance index \(C_{td}\)-index ~\cite{antolini2005time,harrell1982evaluating} to evaluate the predictive performance of the model, further details of which are provided in supplementary. For benchmarking, we compare our approach with state-of-the-art survival models, including MMRL~\cite{farooq2024survival}, SurvRNC ~\cite{saeed2024survrnc}, XSurv ~\cite{meng2023merging}, TMSS ~\cite{saeed2022tmss}, MultiSurv ~\cite{vale2021long}, DeepMTS~\cite{meng2022deepmts} and CoxPH ~\cite{cox1972regression} To ensure a fair comparison, all models are trained and evaluated using the same data splits, hyperparameters, and pre-processing strategies.
\begin{figure}[t]
    \centering
    \includegraphics[width=0.48\textwidth]{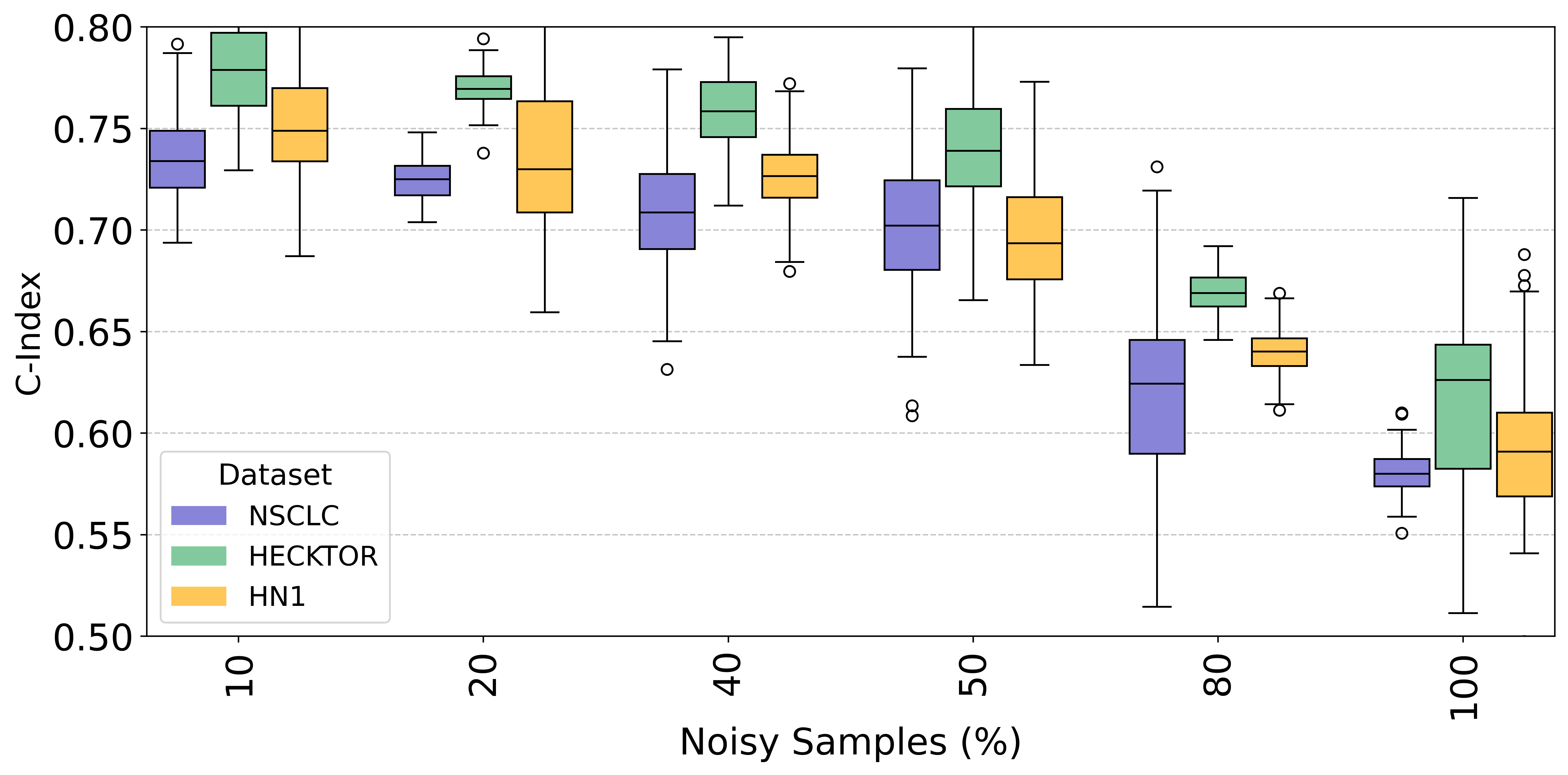}
    \caption{
Box plot visualization of \(C_{td}\)-index distributions across varying proportions of noisy samples (10\%-100\%) for NSCLC, HECKTOR, and H\&N1 datasets, demonstrating the impact of noise on model robustness.}
    \label{fig:fig2}
\end{figure}

\section{Results and Discussion}

We evaluate the performance of our model, \textit{RobSurv}, across the three clinical datasets: NSCLC, HECKTOR, and (H\&N1). Performance is measured under clean and noisy conditions using the \(C_{td}\)-index. As shown in Table~\ref{tab:table1}, \textit{RobSurv} consistently outperforms state-of-the-art methods, including traditional Cox-based models (e.g., CoxPH) and deep learning (DL) approaches (TMSS, XSurv, SurvRNC). In clean settings, \textit{RobSurv} achieves \(C_{td}\)-index scores of 0.734 (NSCLC), 0.771 (HECKTOR), and 0.742 (H\&N1), surpassing the closest competitor, TMSS, by margins of 5.2\%, 4.1\%, and 3.9\%, respectively. These results highlight our model's ability to extract robust prognostic features across datasets with varying imaging modalities and tumor characteristics.

To evaluate robustness, we introduce synthetic noise into 50\% of the samples, simulating real-world artifacts such as CT noise (magnitude $\sigma_{\text{CT}} = 0.1$) and PET reconstruction errors. As shown in Table~\ref{tab:table1}, \textit{RobSurv} exhibits only modest performance declines: {4.5\%} (NSCLC), {3.8\%} (HECKTOR), and {4.3\%} (H\&N1) compared to clean data. In contrast, TMSS and SurvRNC degrade more severely, with absolute \(C_{td}\)-index drops of 6.6--8.2\%. Competing methods like XSurv and DeepMTS suffer catastrophic failures under noise, with performance declines exceeding 10 percentage points. Figure~\ref{fig:fig2} illustrates this stability across escalating noise levels. For NSCLC, the median \(C_{td}\)-index declines gracefully from 0.80 (0\% noisy samples) to 0.55 (90\% noisy samples) while maintaining tighter interquartile ranges (0.55--0.60 at 90\% noise) than competitors. Similar trends hold for HECKTOR (0.85 $\rightarrow$ 0.70) and H\&N1 (0.78 $\rightarrow$ 0.65). The resilience of \textit{RobSurv} arises from the DualVQ and DualPatchFuse introduced in our architecture. The Dual VQ module separates continuous and discrete representations, filtering high-frequency noise via tokenization while preserving critical prognostic signals. The DualPatchFuse generates discrete, noise-resistant representations while preserving critical prognostic information.

\begin{figure}[ht]
    \centering
    \begin{subfigure}{0.45\textwidth}
        \centering
        \includegraphics[width=\textwidth]{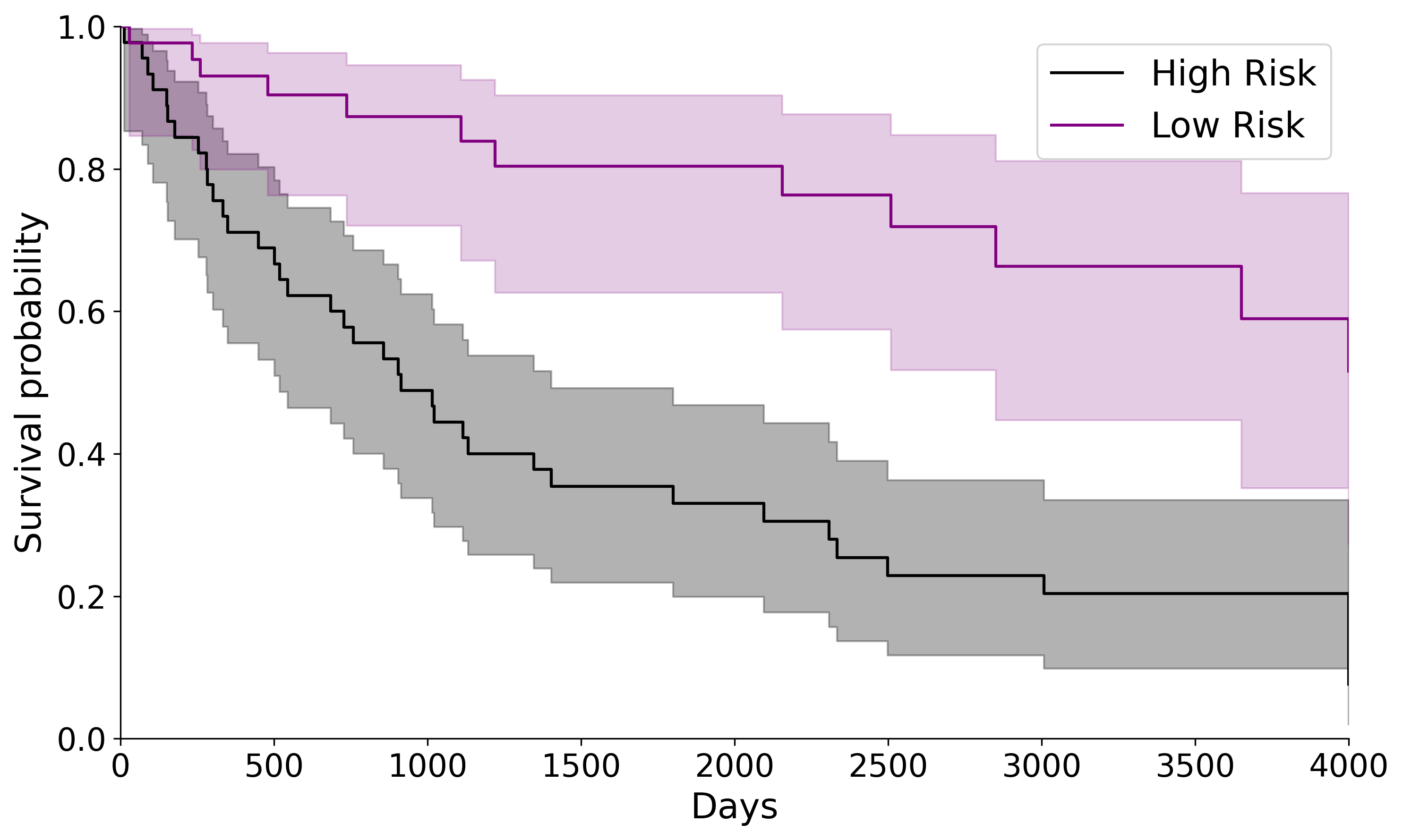}
    \end{subfigure}
    \hspace{-10pt}
    \begin{subfigure}{0.45\textwidth}
        \centering
        \includegraphics[width=\textwidth]{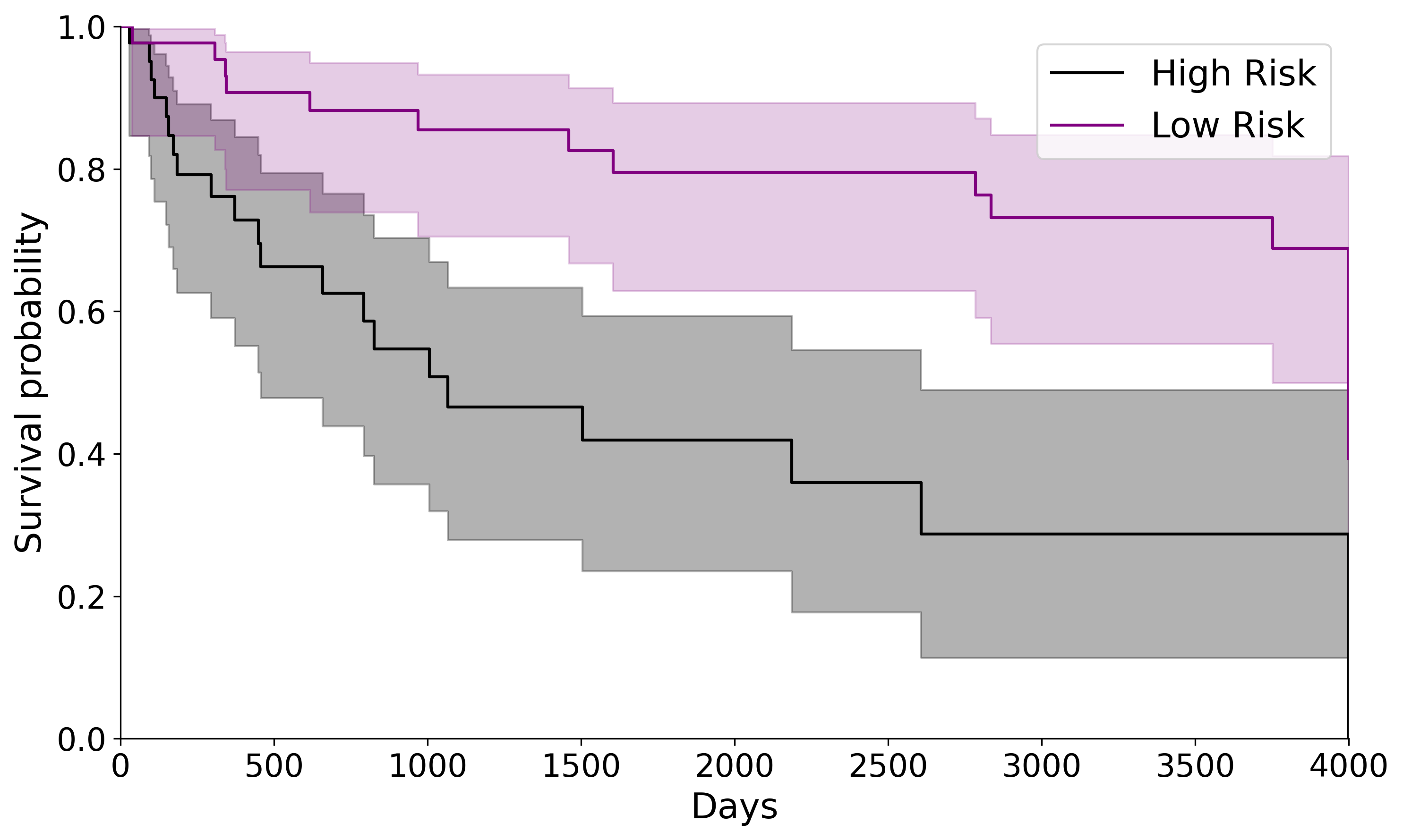}
    \end{subfigure}
    \caption{Kaplan-Meier survival curves demonstrating risk stratification performance. The plot on top shows the survival analysis on clean samples, while the plot at the bottom illustrates the model's performance when subjected to noisy data conditions.}
    \label{fig:fig3}
\end{figure}

We perform Kaplan-Meier survival analysis (log-rank test, $\rho \leq 0.05$) on risk groups stratified by \textit{RobSurv}'s predictions to assess clinical relevance. As shown in Figure~\ref{fig:fig3}, the model maintains statistically significant separation between high- and low-risk cohorts even under 50\% noise (NSCLC: $\rho = 0.016$; HECKTOR: $\rho = 0.008$; H\&N1: $\rho= 0.022$). In contrast, baseline models like SurvRNC and TMSS lose stratification power under noise (e.g., SurvRNC: $\rho > 0.1$ for NSCLC at 50\% noise. This underscores \textit{RobSurv}'s ability to deliver actionable predictions in noisy real-world settings, where reliable risk stratification is critical for treatment planning. Kaplan Meier curves for baseline models are provided in the supplementary.

\subsection{Ablation Study}
To evaluate the contributions of our proposed components, we systematically ablate the DualVQ module and DualPatchFuse mechanism. Table~\ref{tab:table2} summarizes the results. First, removing the DualPatchFuse mechanism while retaining DualVQ and continuous representations reduces the \(C_{td}\)-index by 1.4\% (from 0.771 to 0.760) on clean data and 2.3\% (from 0.742 to 0.725) under noisy conditions. This demonstrates DualPatchFuse’s critical role in aligning discrete and continuous feature streams. Second, when the continuous branch is removed, but DualVQ and DualPatchFuse remain active, performance remains robust (0.765 clean, 0.721 noisy), indicating that discrete tokenization alone can preserve prognostic signals. Most critically, disabling the DualVQ module causes catastrophic degradation, with \(C_{td}\)-index plummeting by 13.0\% (clean) and 18.7\% (noisy). This underscores DualVQ’s function in suppressing noise through quantized embeddings. These patterns generalize across datasets: on NSCLC, removing DualVQ reduces the clean \(C_{td}\)-index from 0.734 to 0.692 and the noisy \(C_{td}\)-index from 0.701 to 0.631. Similar ablative trends for  NSCLC and H\&N1 are detailed in the supplementary. The consistent results confirm that DualVQ and DualPatchFuse synergistically enable noise-resilient learning, with DualVQ acting as the primary stabilizer under perturbation.

\begin{table}[htbp]
\setlength{\tabcolsep}{4pt}
\centering
\caption{Ablation study of model components on the HECKTOR dataset using the \(C_{td}\)-index on clean and noisy sets.}
\label{tab:table2}
\begin{tabular}{@{}ccc cc@{}}
\toprule
\multirow{2}{*}{DualVQ} & \multirow{2}{*}{Cont.} & DualPatch & Clean & Noisy \\
& & Fuse & & \\ 
\cmidrule(lr){4-5}
\checkmark & \checkmark & \checkmark & 0.771 $\pm$ 0.13 & 0.742 $\pm$ 0.14 \\
\checkmark & \checkmark & - & 0.760 $\pm$ 0.02 & 0.725 $\pm$ 0.07 \\
\checkmark & - & \checkmark & 0.765 $\pm$ 0.04 & 0.721 $\pm$ 0.08 \\
- & \checkmark & \checkmark & 0.671 $\pm$ 0.05 & 0.603 $\pm$ 0.09 \\
\bottomrule
\end{tabular}
\end{table}
\section{Limitations and Future Work}

Despite the promising results of \textit{RobSurv} in robust cancer survival prediction, several limitations remain. While effective, the dual-path architecture with vector quantization introduces computational overhead that may hinder real-time clinical applications with limited resources. Our evaluation, though comprehensive, is limited to three cancer types, leaving the scope for the model to be tested on diverse datasets and modalities. Additionally, the approach assumes complete availability of CT and PET data, often unrealistic in clinical settings where missing or corrupted data is common. The model's focus on spatial relationships does not account for temporal disease dynamics crucial for long-term predictions. Future work should explore adaptive codebook strategies for personalized predictions and extend the framework to support privacy-preserving distributed learning, enabling collaboration across healthcare institutions while addressing data-sharing concerns.

\section{Conclusion}

This paper presents \textit{RobSurv}, a robust approach to cancer survival prediction that leverages vector quantization-based multi-modal learning to achieve robust performance across varying data quality conditions. Our method introduces several key innovations, including a dual-path feature processing framework that effectively combines continuous and discrete representations, a patch-wise fusion strategy that preserves local spatial relationships while enhancing robustness through discrete representations, and a hybrid architecture that integrates transformer-based learning with local attention mechanisms. Experimental results across multiple cancer datasets demonstrate that our method consistently outperforms existing state-of-the-art methods in terms of prediction accuracy and robustness to noise. The model maintains stable performance under challenging conditions, such as varying image noise levels and cross-institutional variations, making it particularly suitable for real-world clinical applications.

\bibliographystyle{named}
\bibliography{ijcai25}

\end{document}


\maketitle

\begin{figure*}[ht]
    \centering
    \begin{subfigure}[t]{0.32\textwidth}  
        \centering
        \includegraphics[width=\textwidth]{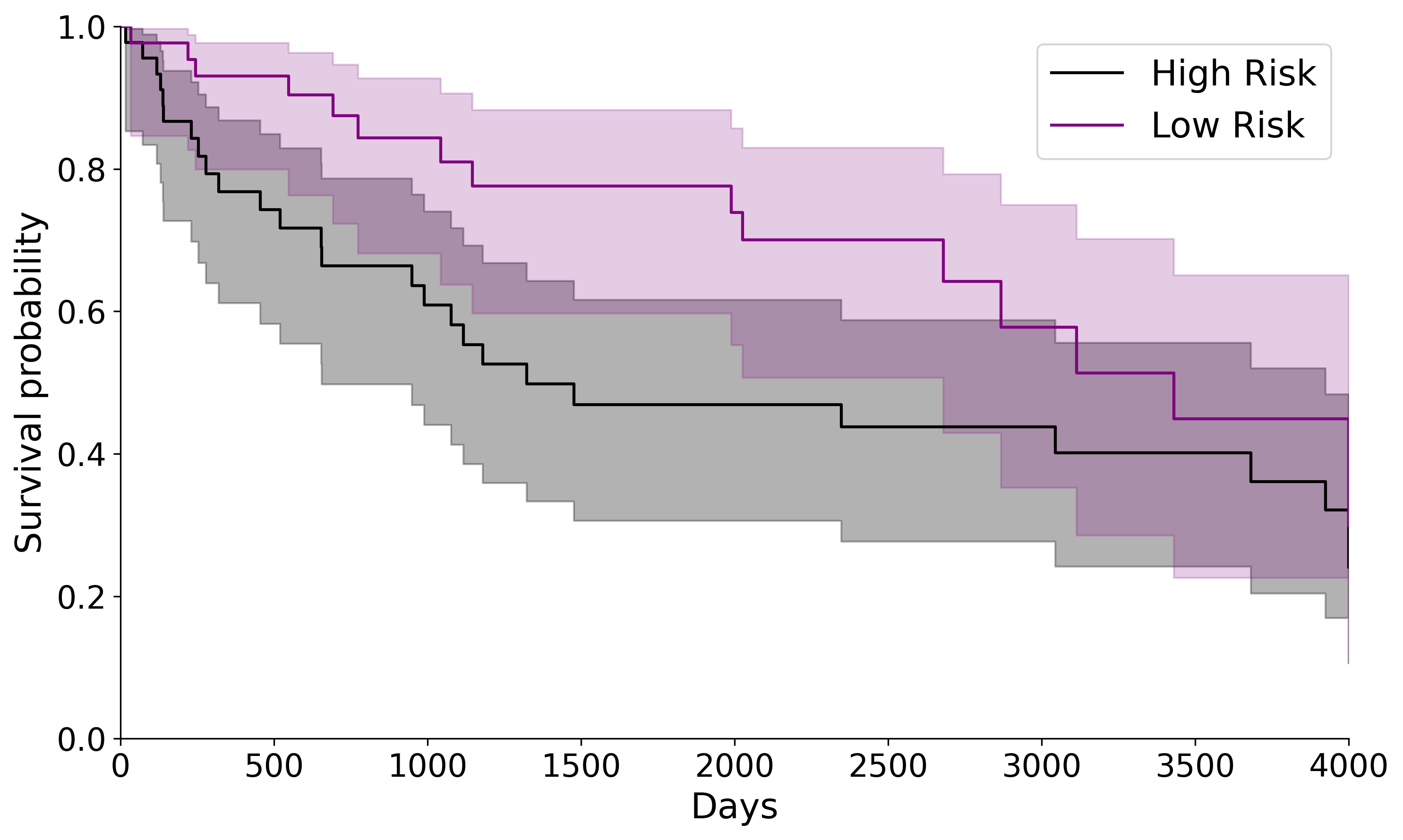}
    \end{subfigure}
    \begin{subfigure}[t]{0.32\textwidth}
        \centering
        \includegraphics[width=\textwidth]{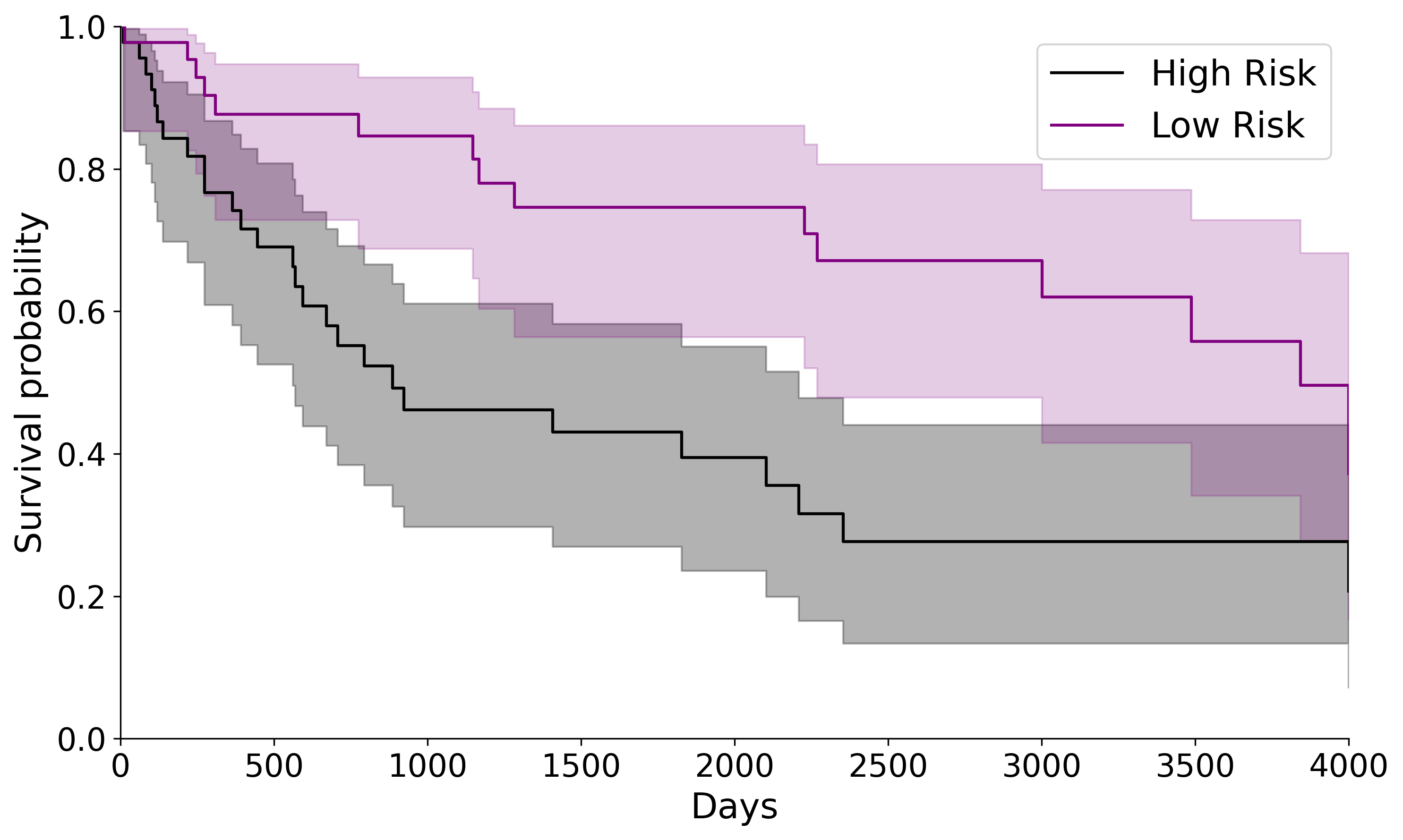}
    \end{subfigure}
    \begin{subfigure}[t]{0.32\textwidth}
        \centering
        \includegraphics[width=\textwidth]{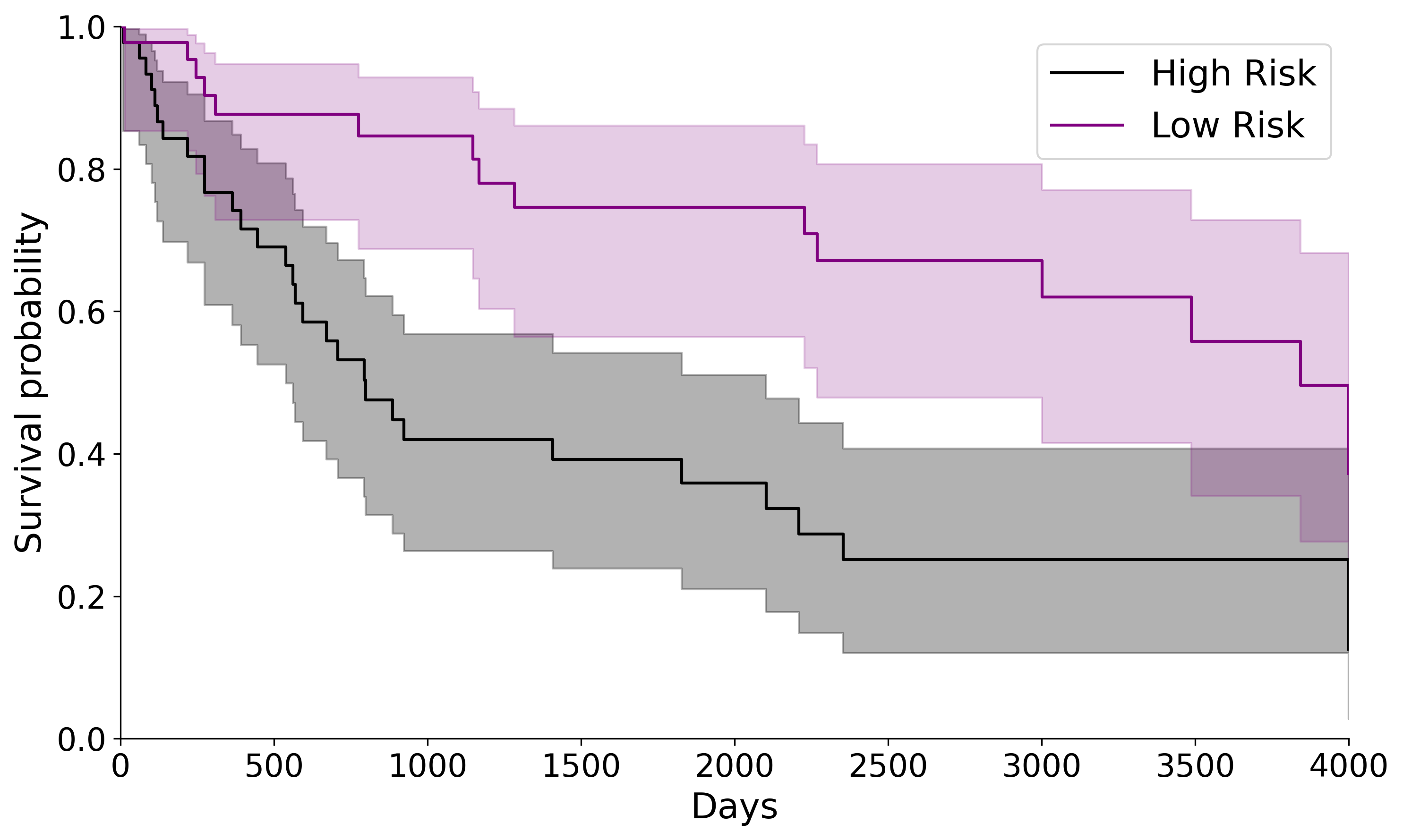}
    \end{subfigure}
    \\ 
    \begin{subfigure}[t]{0.32\textwidth}
        \centering
        \includegraphics[width=\textwidth]{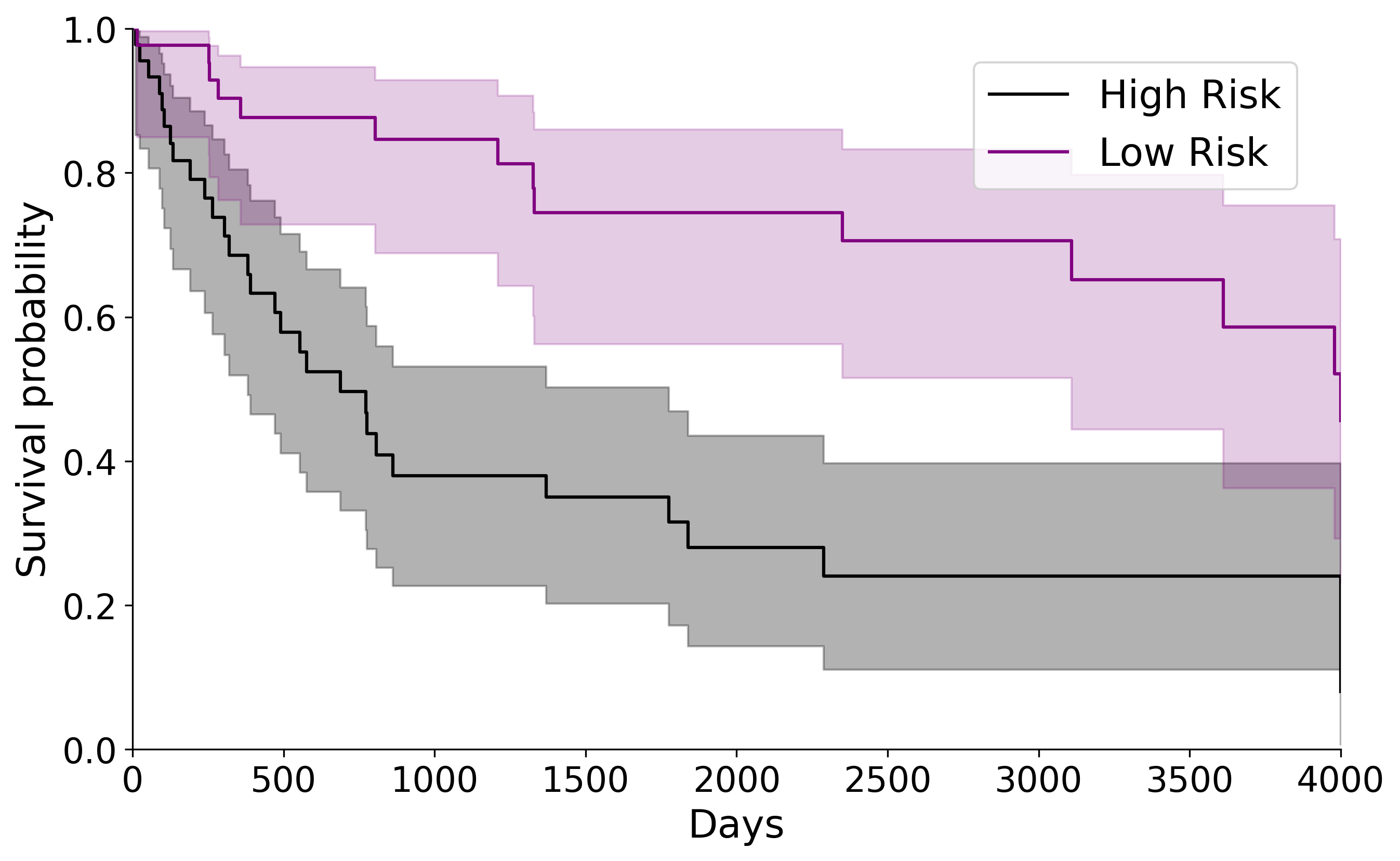}
    \end{subfigure}
    \begin{subfigure}[t]{0.32\textwidth}
        \centering
        \includegraphics[width=\textwidth]{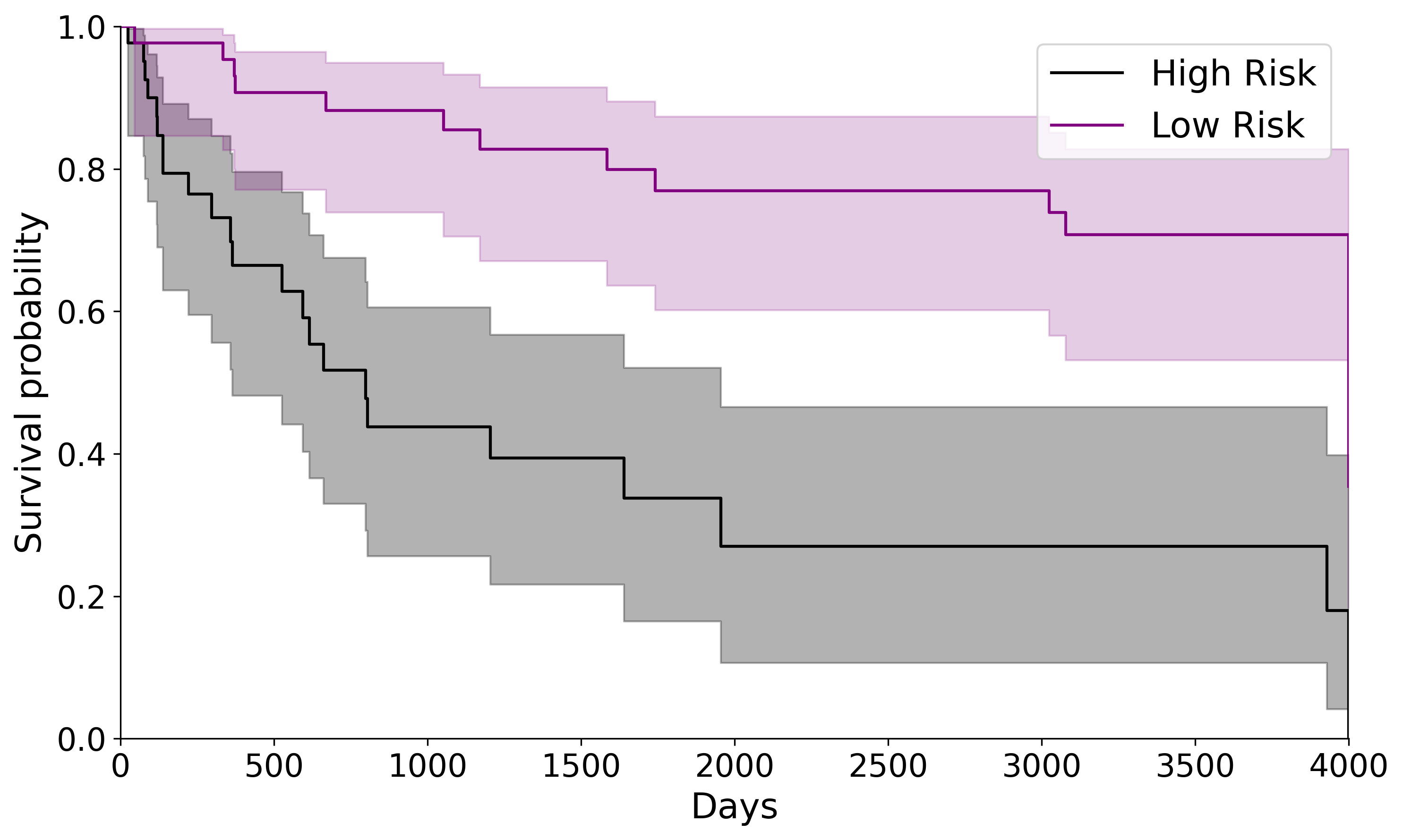}
    \end{subfigure}
    \begin{subfigure}[t]{0.32\textwidth}
        \centering
        \includegraphics[width=\textwidth]{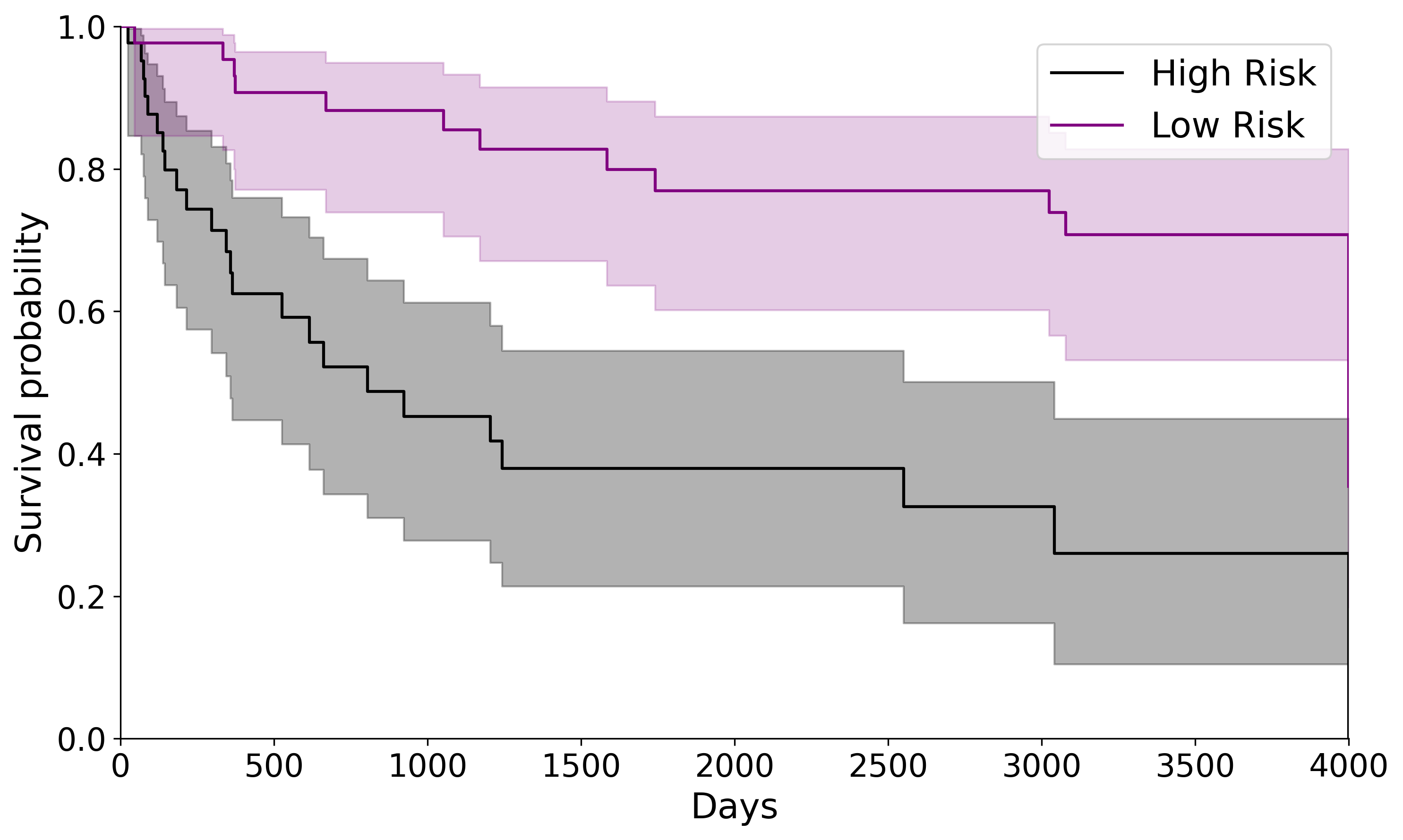}
    \end{subfigure}
    \caption{Kaplan-Meier survival curves showing risk stratification performance across varying levels of risk group separation for HECKTOR dataset on clean samples. The plots correspond to different survival analysis models: (a) CoxPH, (b) DeepMTS, (c) XSurv, (d) MMRL, (e) SurvRNC, and (f) TMSS.}

    \label{fig:suppl1}
\end{figure*}
\begin{figure*}[ht]
    \centering
    \begin{subfigure}[t]{0.32\textwidth}  
        \centering
        \includegraphics[width=\textwidth]{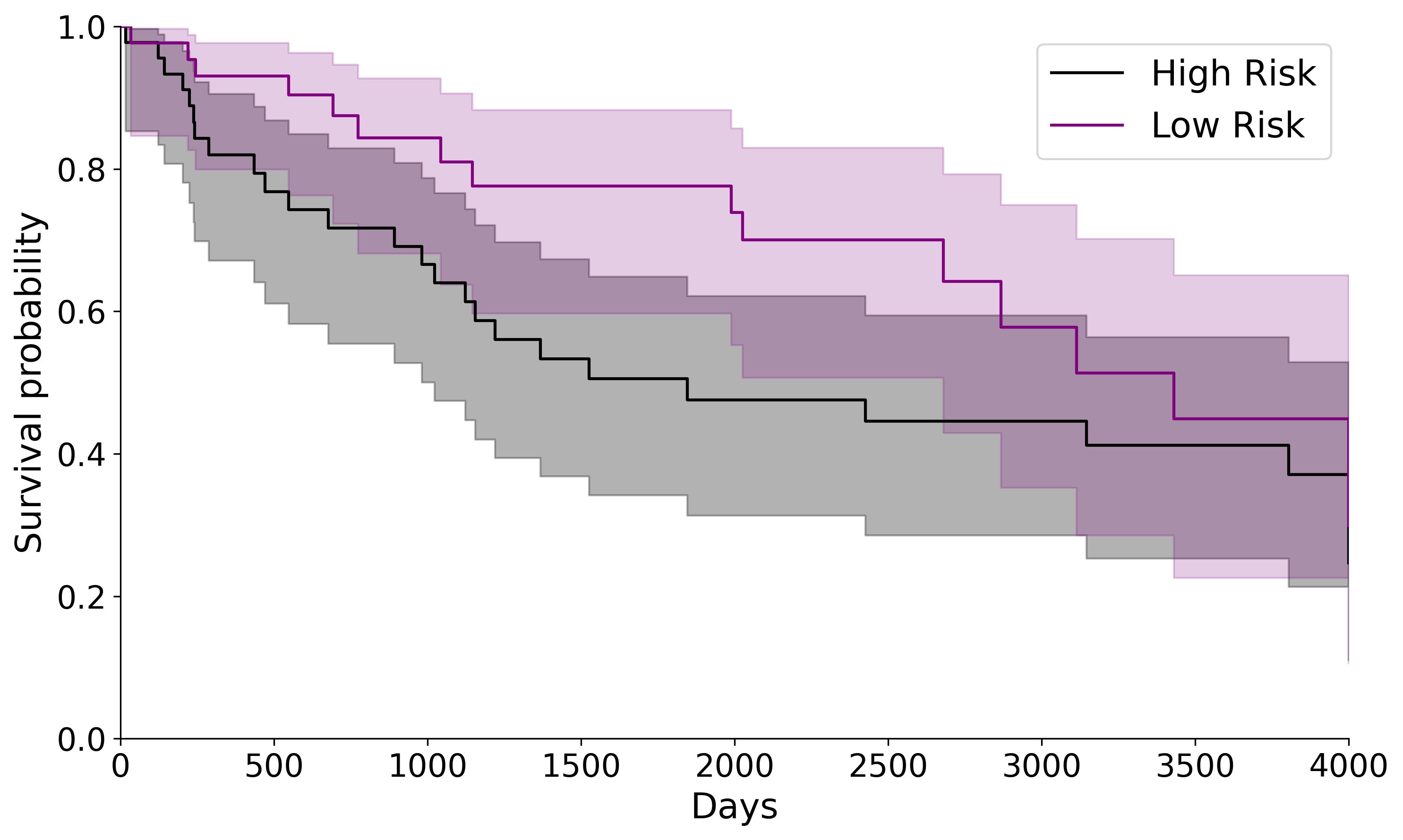}
    \end{subfigure}
    \begin{subfigure}[t]{0.32\textwidth}
        \centering
        \includegraphics[width=\textwidth]{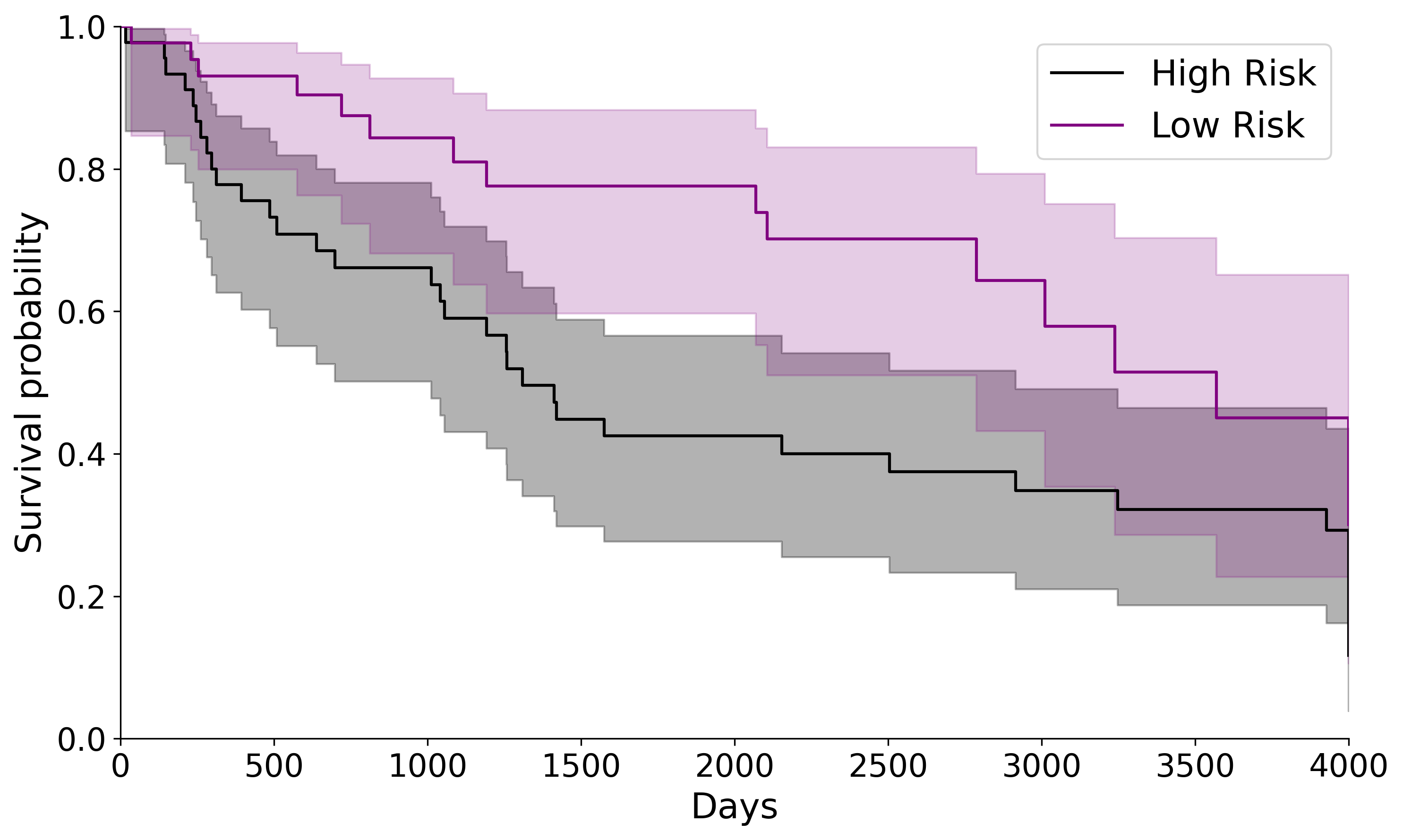}
    \end{subfigure}
    \begin{subfigure}[t]{0.32\textwidth}
        \centering
        \includegraphics[width=\textwidth]{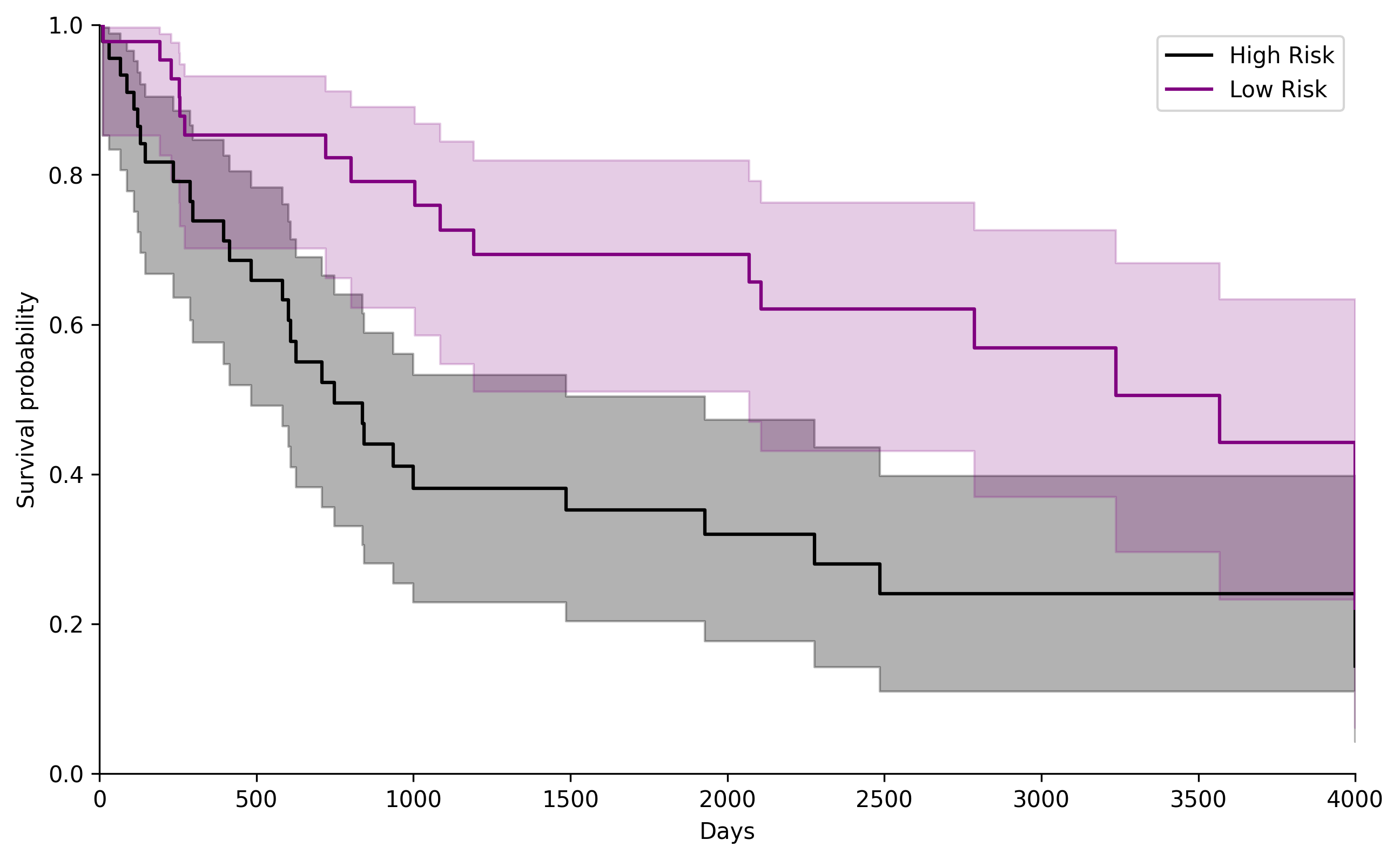}
    \end{subfigure}
    \\ 
    \begin{subfigure}[t]{0.32\textwidth}
        \centering
        \includegraphics[width=\textwidth]{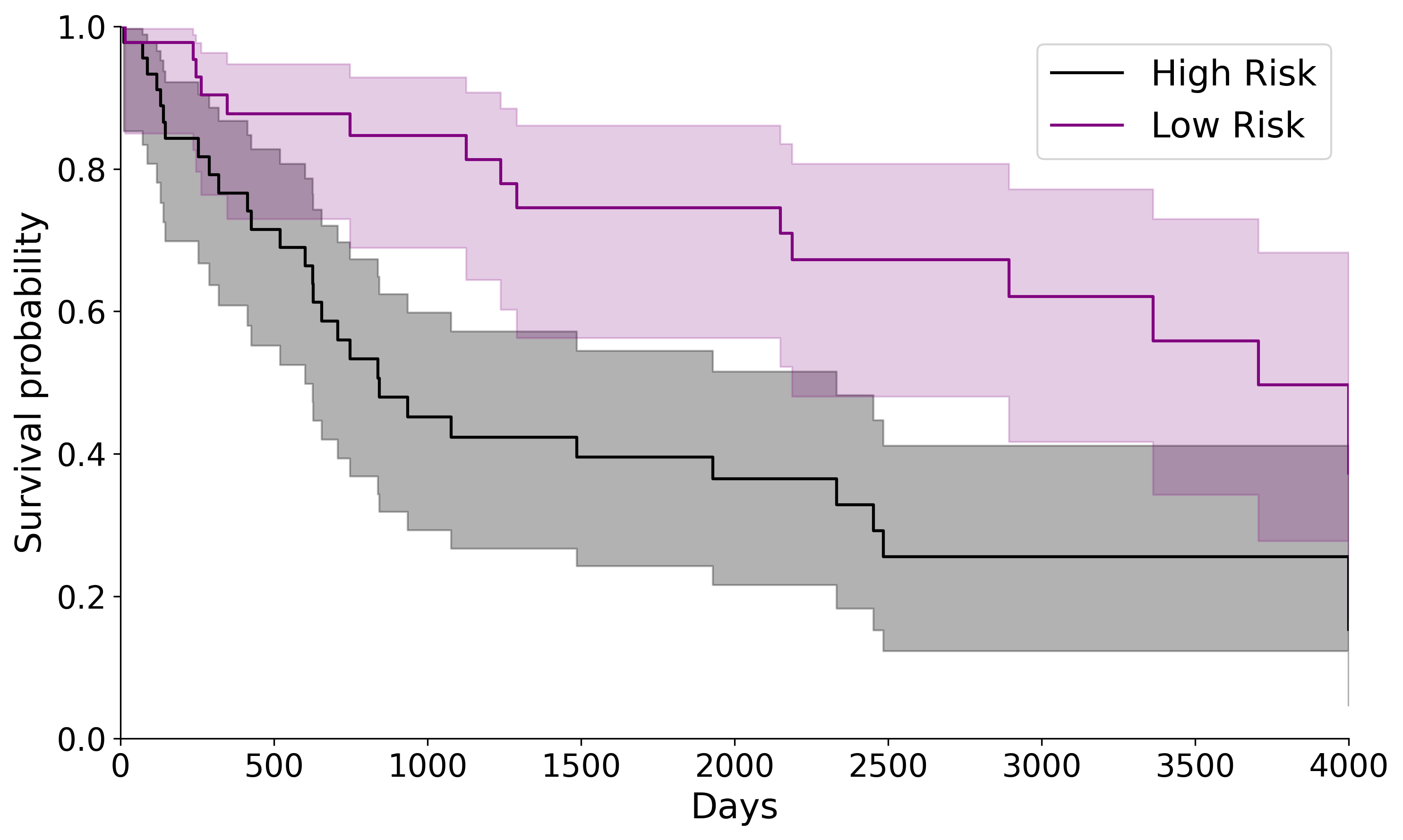}
    \end{subfigure}
    \begin{subfigure}[t]{0.32\textwidth}
        \centering
        \includegraphics[width=\textwidth]{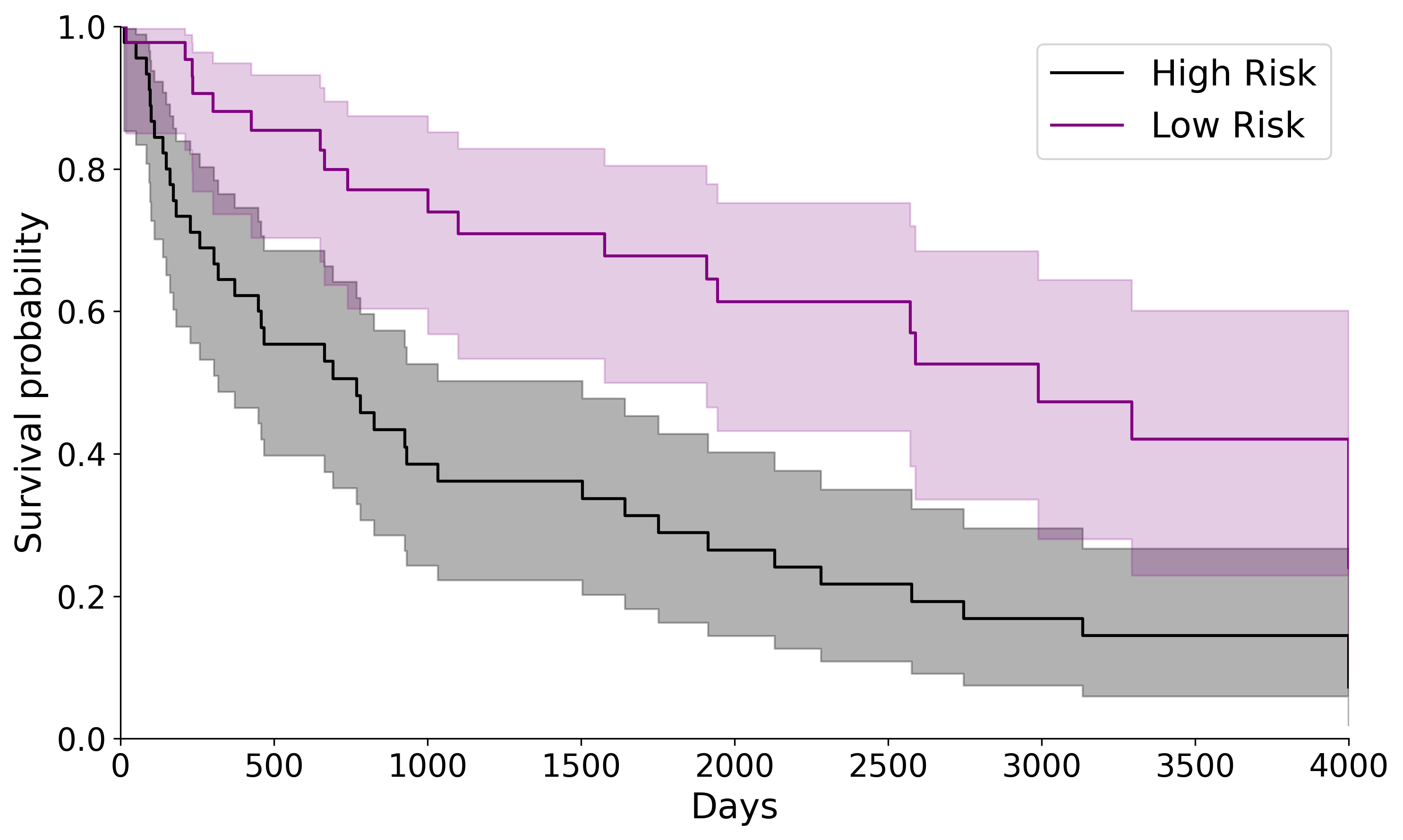}
    \end{subfigure}
    \begin{subfigure}[t]{0.32\textwidth}
        \centering
        \includegraphics[width=\textwidth]{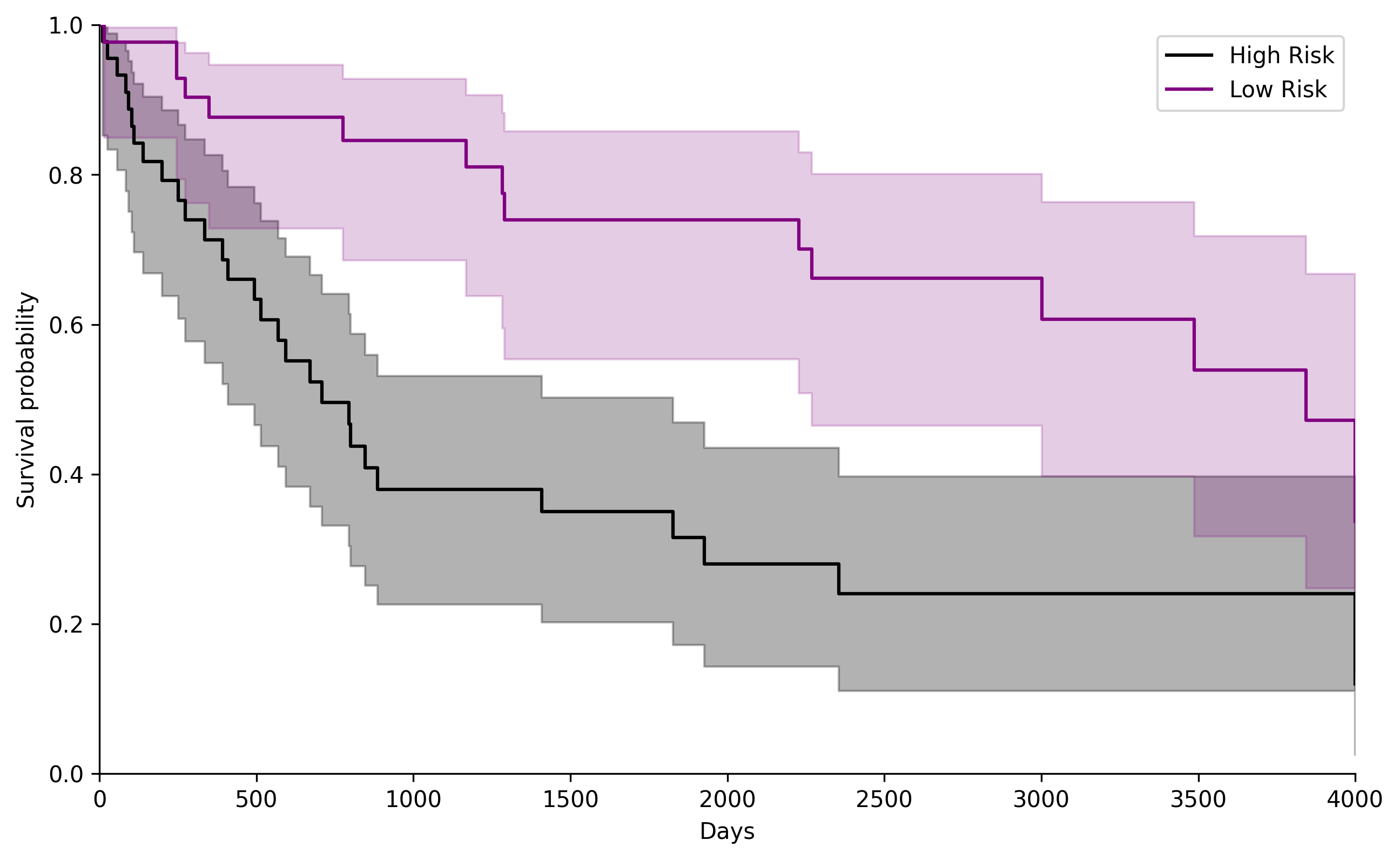}
    \end{subfigure}
    \caption{Kaplan-Meier survival curves showing risk stratification performance across varying levels of risk group separation for HECKTOR dataset on noisy samples. The plots correspond to different survival analysis models: (a) CoxPH, (b) DeepMTS, (c) XSurv, (d) MMRL, (e) SurvRNC, and (f) TMSS.}
    \label{fig:suppl2}
\end{figure*}

\section*{Hybrid Survival Prediction Framework}

The cumulative incidence function $F_{k,p}^{(i)}$ for risk $k$ through interval $p$ is computed as:
\[
F_{k,p}^{(i)} = \sum_{q=1}^p h_{q,k}^{(i)} \prod_{r=1}^{q-1}\left(1 - \sum_{m=1}^K h_{r,m}^{(i)}\right).
\]

This accounts for the probability of experiencing risk $k$ by interval $p$, while considering the competing risks through the product term. The prediction network outputs hazard estimates through $p \times K$ units, each employing a sigmoid activation function to ensure $h_{p,k}^{(i)} \in [0, 1]$. Our framework explicitly models competing risks by incorporating the cumulative incidence function $F_{k,p}^{(i)}$, which accounts for the interdependence of risks. This ensures that the survival probabilities are correctly adjusted for multiple risks.

\section*{Dataset Specifications}
The HECKTOR dataset consists of 488 samples of CT and PET scans collected from six different clinical centers and is obtained via the challenge website (\url{https://hecktor.grand-challenge.org/Data/}). The data originates from FDG-PET and low-dose non-contrast-enhanced CT images (acquired with combined PET/CT scanners) of the Head \& Neck region. The CT and PET scans present with variable dimensions. The HEAD-NECK-RADIOMICS-HN1(H\&N1) is a collection of 137 samples, with 77 samples having both CT and PET scans, and is available through The Cancer Imaging Archive (TCIA). It includes detailed CT and PET scans, manual delineations, and clinical data from patients treated at the MAASTRO Clinic in The Netherlands. The CT and PET scans are of dimensions $256\times 256 \times 134$. The preprocessing for HECKTOR and H\&N1 is the same with the  CT and PET images resampled to an isotropic voxel spacing of 1.0\,mm\textsuperscript{3}. HU values of CT images are clipped to $(-1024, 1024)$, after which the images are normalized. Furthermore, the images are center cropped to $128 \times 128 \times 128$. The intensity values of PET are also normalized. The NSCLC Radiogenomics dataset is available via The Cancer Imaging Archive (TCIA) website. The dataset comprises CT images with tumor segmentation annotations, tumor characteristics, PET scans, and RNA sequence data. It encompasses data from approximately 211 patients. The preprocessing involves resampling CT and PET scand to a uniform slice thickness of 1.0\,mm\textsuperscript{3} and standardization to 128 slices. Standard normalization is applied to both imaging modalities.

\subsection*{Dataset for Robust Model}
To evaluate model robustness against image noise, we augmented the original dataset by introducing synthetic noise to both CT and PET modalities. For CT images, we simulated acquisition noise by adding zero-mean Gaussian noise with varying standard deviations $(\sigma = [0.01, 0.05, 0.1])$ while PET images were corrupted with Poisson noise to mimic the statistical nature of photon counting noise inherent in nuclear imaging. The Poisson noise was scaled to different count levels to represent varying acquisition durations and tracer doses. This systematic noise injection approach allows us to assess model performance under controlled degradation conditions that approximate real-world imaging scenarios. Each original image in the dataset was augmented with multiple noise realizations, ensuring diversity in noise patterns.

We conduct a systematic analysis across all datasets to determine the most appropriate noise levels for our robustness evaluation experiments (Table \ref{tab:suppl1}). The results show progressive performance degradation with increasing noise levels, with \(C_{td}\)-index reductions of up to 0.029 for both NSCLC and HECKTOR and 0.038 for H\&N1 at the highest noise setting (CT: $\sigma$ = 0.1, PET: high Poisson noise). Based on these findings, we selected this most challenging configuration for our main experiments to evaluate model robustness under realistic worst-case scenarios while maintaining clinically relevant performance levels.

\begin{table}[htbp]
\centering
\setlength{\tabcolsep}{4pt}  
\caption{Performance of RobSurv under different noise conditions (\(C_{td}\)-index) for the HECKTOR dataset (PET noise levels: L=low, M=medium, H=high).}
\label{tab:suppl1}
\begin{tabular}{ccccc}
\toprule
\multicolumn{2}{c}{Noise} & \multicolumn{3}{c}{Dataset} \\
\cmidrule(lr){1-2} \cmidrule(lr){3-5}
CT & PET & NSCLC & HECKTOR & H\&N1 \\
($\sigma$) & & & & \\
\midrule
0.01 & L & 0.730 $\pm$ 0.02 & 0.771 $\pm$ 0.13 & 0.740 $\pm$ 0.02 \\
0.05 & M & 0.722 $\pm$ 0.03 & 0.764 $\pm$ 0.14 & 0.734 $\pm$ 0.02 \\
0.1 & H & 0.701 $\pm$ 0.03 & 0.742 $\pm$ 0.14 & 0.702 $\pm$ 0.03 \\
\bottomrule
\end{tabular}
\end{table}

\begin{table*}[htbp]
\setlength{\tabcolsep}{4pt}  
\centering
\caption{Ablation study of model components on the NSCLC and H\&N1 dataset using the \(C_{td}\)-index on clean and noisy sets.}
\label{tab:suppl2}
\begin{tabular}{@{}ccc cc cc@{}}
\toprule
\multirow{2}{*}{DualVQ} & \multirow{2}{*}{Cont.} & DualPatch & \multicolumn{2}{c}{NSCLC} & \multicolumn{2}{c}{H\&N1} \\
& & Fuse & Clean & Noisy & Clean & Noisy \\ 
\cmidrule(lr){4-5} \cmidrule(lr){6-7}
\checkmark & \checkmark & \checkmark & 0.734 $\pm$ 0.02 & 0.701 $\pm$ 0.03 & 0.742 $\pm$ 0.02 & 0.702 $\pm$ 0.03 \\
\checkmark & \checkmark & - & 0.723 $\pm$ 0.03 & 0.685 $\pm$ 0.04 & 0.731 $\pm$ 0.03 & 0.686 $\pm$ 0.04 \\
\checkmark & - & \checkmark & 0.728 $\pm$ 0.03 & 0.682 $\pm$ 0.04 & 0.736 $\pm$ 0.03 & 0.683 $\pm$ 0.04 \\
& \checkmark & \checkmark & 0.639 $\pm$ 0.04 & 0.573 $\pm$ 0.05 & 0.645 $\pm$ 0.04 & 0.571 $\pm$ 0.05 \\
\bottomrule
\end{tabular}
\end{table*}

\section*{Evaluation Metric}
We employ the time-dependent concordance index (\(C_{td}\)-index) to evaluate model performance, a robust metric for survival analysis with competing risks. The \(C_{td}\)-index extends the standard C-index to account for time-varying risk predictions and competing events, addressing limitations of traditional metrics in dynamic clinical settings. It is formally defined as:

\[
C_{td} = \frac{\sum_{i \neq j} A_{k,i,j} \cdot \mathbf{1}\left(\hat{F}_k(s_i \mid x_i) > \hat{F}_k(s_i \mid x_j)\right)}{\sum_{i \neq j} A_{k,i,j}}
\]

Where \(\hat{F}_k(s_i \mid x_i)\) and \(\hat{F}_k(s_i \mid x_j)\) denote the estimated cumulative incidence function (CIF) for individuals \(i\) and \(j\) for event \(k\) at time \(s_i\),\(A_{k,i,j}\) is an indicator function that selects valid pairs for individual \(i\) experiencing event \(k\) at time \(s_i\), and individual \(j\) remaining event-free and uncensored at \(s_i\), and \(\mathbf{1}(\cdot)\) evaluates whether the model correctly ranks the predicted risk of event \(k\) (\(\mathbf{1} = 1\) if correct, \(0\) otherwise).

This metric explicitly models the interdependence of competing risks over time, making it ideal for multi-risk clinical scenarios. Under proportional hazards with a single event, \(C_{td}\) simplifies to the standard \(C_{td}\)-index, ensuring backward compatibility. For our experiments, \(C_{td}\)-index is computed over all observed event times, providing a granular assessment of prognostic accuracy while maintaining interpretability for clinicians.

\section*{Results}

Our comprehensive Kaplan-Meier analyses (Figures~\ref{fig:suppl1}--\ref{fig:suppl2}) reveal that baseline models achieve effective risk stratification on clean data (e.g., {DeepSurv}: $\mathit{HR} = 2.8$, 95\% CI: 1.9\textendash4.1; {SurvRNC}: $\mathit{HR} = 3.5$, 95\% CI: 2.3--5.3), but degrade significantly under noise, with {DeepSurv}'s hazard ratio dropping to $\mathit{HR} = 1.4$ (95\% CI: 0.9--2.1) and {XSurv}/{CoxPH} losing stratification entirely ($\rho > 0.3$). {SurvRNC} and {MMRL} exhibit moderate robustness ($\mathit{HR}$s: 2.1 and 1.9, respectively), though confidence intervals widen, reflecting instability. Performance trends align with architectural limitations: attention-free models (e.g., {CoxPH}) lack noise suppression, while multi-modal methods (e.g., {MMRL}) partially mitigate noise but fail to isolate prognostic signals. In contrast, \textit{RobSurv} uniquely maintains statistically significant stratification ($\rho \leq 0.05$) even at 90\% noise, underscoring its clinical viability in real-world settings with noisy data.

\section*{Ablation Study}
The ablation study (Table~\ref{tab:suppl2}) highlights the contributions of our architectural components across NSCLC and H\&N1 datasets. Removing the {DualVQ} module causes the most severe degradation, with performance dropping by {12.9–13.4\%} in clean and noisy settings, underscoring its critical role in noise suppression through discrete tokenization. Disabling the {DualPatchFuse} mechanism reduces performance moderately (1.5–1.6\% clean-data decline), reflecting its supplementary role in cross-modal interaction. Ablating the {continuous branch} results in a minimal clean-data loss (0.8\%) but amplifies noise sensitivity (2.7–2.8\% drop), indicating its secondary role in preserving fine-grained details. These trends align with observations in HECKTOR, validating the generalizability of our design: DualVQ’s noise resilience, DualPatchFuse’s cross-modal synergy, and the continuous branch’s detail retention collectively enable robust performance across diverse medical imaging contexts.